\documentclass[preprint2]{aastex}

\slugcomment{Received...;Accepted}

\shorttitle{The complex physics of dusty star-forming galaxies at high redshifts}
\shortauthors{Lo Faro et al.}

\usepackage{graphicx}
\usepackage{multicol}
\usepackage{lscape}
\usepackage{overpic,color}
\usepackage{array}

\begin{document}

\title{The complex physics of dusty star-forming galaxies at high redshifts as revealed by Herschel and Spitzer 
\thanks{\textit{Herschel} is an ESA space observatory with science instruments provided by European-led Principal Investigator consortia and with important participation from NASA.}}

\author{B.~Lo~Faro\altaffilmark{1,2}, A.~Franceschini\altaffilmark{1}, M.~Vaccari\altaffilmark{1,18}, L.~Silva\altaffilmark{3}, G.~Rodighiero\altaffilmark{1}, S.~Berta\altaffilmark{4}, J.~Bock\altaffilmark{6,7}, D.~Burgarella\altaffilmark{10}, V.~Buat\altaffilmark{10}, A.~Cava\altaffilmark{19}, D.L.~Clements\altaffilmark{17}, A.~Cooray\altaffilmark{11}, D.~Farrah\altaffilmark{5,20}, A.~Feltre\altaffilmark{1}, E.A.~Gonz\'alez~Solares\altaffilmark{8}, P.~Hurley\altaffilmark{5}, D.~Lutz\altaffilmark{4}, G.~Magdis\altaffilmark{13}, B.~Magnelli\altaffilmark{4}, L.~Marchetti\altaffilmark{1}, S.J.~Oliver\altaffilmark{5}, M.J.~Page\altaffilmark{14}, P.~Popesso\altaffilmark{4}, F.~Pozzi\altaffilmark{4}, D.~Rigopoulou\altaffilmark{15,16}, M.~Rowan-Robinson\altaffilmark{17}, I.G.~Roseboom\altaffilmark{5,9}, Douglas~Scott\altaffilmark{12}, A.J.~Smith\altaffilmark{5}, M.~Symeonidis\altaffilmark{14}, L.~Wang\altaffilmark{5}, S.~Wuyts\altaffilmark{4}}
\altaffiltext{1}{Dipartimento di Fisica e Astronomia, Universit${\grave{a}}$ di Padova, vicolo Osservatorio, 3, 35122 Padova, Italy}
\altaffiltext{2}{barbara.lofaro@studenti.unipd.it}
\altaffiltext{3}{INAF-OATs, Via Tiepolo 11, I-34131 Trieste, Italy}
\altaffiltext{4}{MPE, Postfach 1312, 85741, Garching, Germany}
\altaffiltext{5}{Astronomy Centre, Dept. of Physics \& Astronomy, University of Sussex, Brighton BN1 9QH, UK}
\altaffiltext{6}{California Institute of Technology, 1200 E. California Blvd., Pasadena, CA 91125}
\altaffiltext{7}{Jet Propulsion Laboratory, 4800 Oak Grove Drive, Pasadena, CA 91109}
\altaffiltext{8}{Institute of Astronomy, University of Cambridge, Madingley Road, Cambridge CB3 0HA, UK}
\altaffiltext{9}{Institute for Astronomy, University of Edinburgh, Royal Observatory, Blackford Hill, Edinburgh EH9 3HJ, UK}
\altaffiltext{10}{(Laboratoire d'Astrophysique de Marseille, OAMP, Universit\'{e} Aix-Marseille, CNRS, 38 rue Fr\'{e}d\'{e}ric Joliot-Curie, 13388 Marseille Cedex 13, France}
\altaffiltext{11}{Department of Physics \& Astronomy, University of California, Irvine, CA 92697}
\altaffiltext{12}{Department of Physics \& Astronomy, University of British Columbia, 6224 Agricultural Road, Vancouver, BC V6T~1Z1, Canada}
\altaffiltext{13}{Department of Physics, University of Oxford, Keble Road, Oxford OX1 3RH, UK}
\altaffiltext{14}{Mullard Space Science Laboratory, University College London, Holmbury St. Mary, Dorking, Surrey RH5 6NT, UK}
\altaffiltext{15}{RAL Space, Rutherford Appleton Laboratory, Chilton, Didcot, Oxfordshire OX11 0QX, UK}
\altaffiltext{16}{Department of Astrophysics, Denys Wilkinson Building, University of Oxford, Keble Road, Oxford OX1 3RH, UK}
\altaffiltext{17}{Astrophysics Group, Imperial College London, Blackett Laboratory, Prince Consort Road, London SW7 2AZ, UK}
\altaffiltext{18}{Astrophysics Group, Physics Department, University of the Western Cape, Private Bag X17, 7535, Bellville, Cape Town, South Africa}
\altaffiltext{19}{Departamento de Astrof\'isica, Facultad de CC. F\'isicas, Universidad Complutense de Madrid, E-28040 Madrid, Spain}
\altaffiltext{20}{Department of Physics, Virginia Tech, Blacksburg, VA 24061, USA}

\begin{abstract}
We combine far-infrared photometry from {\textit{Herschel}} (PEP/HERMES) with deep mid-infrared spectroscopy from {\textit{Spitzer}} to investigate the nature and the mass assembly history of a sample of 31 Luminous and Ultraluminous Infrared Galaxies at z$\sim$1 and 2 selected in GOODS-S with $24$ $\mu$m fluxes between 0.2 and 0.5 mJy. We model the data with a self-consistent physical model (GRASIL) which includes a state-of-the-art treatment of dust extinction and reprocessing. We find that all of our galaxies appear to require massive populations of old ($>1$\ Gyr) stars and, at the same time, to host a moderate ongoing activity of SF (SFR $\leq$ 100\ M$_{\odot}$/yr). The bulk of the stars appear to have been formed a few Gyr before the observation in essentially all cases. Only five galaxies of the sample require a recent starburst superimposed on a quiescent star formation history (SFH). We also find discrepancies between our results and those based on optical-only SED fitting for the same objects; by fitting their observed Spectral Energy Distributions with our physical model we find higher extinctions (by $\Delta$A$_{\mathrm{V}}\sim$ 0.81 and 1.14) and higher stellar masses (by $\Delta$Log(M$_{\star})\sim$ 0.16 and 0.36 dex) for z$\sim$1 and z$\sim$2 (U)LIRGs, respectively. The stellar mass difference is larger for the most dust obscured objects. We also find lower SFRs than those computed from $L_{\mathrm{IR}}$ using the Kennicutt relation due to the significant contribution to the dust heating by intermediate-age stellar populations through `cirrus' emission ($\sim$73\% and $\sim$66\% of total $L_{\mathrm{IR}}$ for $z\sim1$ and $z\sim2$ (U)LIRGs, respectively). 
\end{abstract}


\section{Introduction}
\label{introduction}

The spectral energy distribution (SED) of a galaxy contains valuable information about its physical properties, including 
the stellar, gas and dust content, the age and abundance distribution of the stellar populations resulting from the star formation history (SFH), and their interaction with the interstellar medium (ISM). 
The study of the SED therefore offers the most direct way to investigate galaxy formation and evolution, 
both through direct observations and corresponding theoretical modelling. The different physical processes occurring in galaxies all leave their imprint on the global and detailed shape of the spectrum, each dominating at different wavelengths. Therefore, by analysing and predicting the whole spectral range one can hope to deconvolve and
interpret all the information contained in the SED in terms
of the SFH and galaxy evolution in general. 

Stellar sources mainly emit in the UV/optical to NIR spectral range, and the SED in this wavelength region is therefore heavily influenced by the star
formation history of the galaxy. The SED from a few $\mu$m to the sub-mm (the IR region) is dominated by the interaction of dust grains with stellar radiation. Basically dust absorbs and scatters photons, mostly at wavelength $\lambda \lesssim$  1 $\mu$m and thermally emit the absorbed energy in the IR. The resulting SED is often substantially changed and in many relevant cases radically modified by the presence of dust.

A detailed modelling of the entire SED of galaxies is therefore very complex. Several different approaches have been proposed, depending on the purpose of the applications. Some works (e.g. Devriendt, Guiderdoni \& Sadat~1999; Chary \& Elbaz 2001; Dale et al.~2001; Dale \& Helou~2002; Galliano et al.~2003; Lagache, Dole \& Puget~2003; Burgarella et al.~2005, da Cunha, Charlot \& Elbaz~2008) have proposed semi-empirical treatments of the SEDs. The aim of these works is in general to interpret large samples of data, requiring fast computing times making use of observationally or physically motivated SEDs. There are then other codes, as Hyperz (Bolzonella et al.~2000) or LePhare (Arnouts et al.~1999) which usually perform a template-based SED fitting procedure based on a standard $\chi^{2}$ minimization method. Usually here dust effects are accounted for by computing the attenuation of the light from a source placed behind a foreground screen of dust and assuming different extinction curves. Other works are based, instead, on theoretical computations in order to have a more general applicability in terms of interpretative and predictive power. Several papers deal with the radiative transfer (RT) in spherical geometries, mainly aimed at modelling starburst galaxies (e.g. Rowan-Robinson~1980; Rowan-Robinson \& Crawford~1989; Efstathiou, Rowan-Robinson \& Siebenmorgen~2000; Popescu et al.~2000; Efstathiou \& Rowan-Robinson~2003; Takagi, Arimoto \& Hanami~2003a; Takagi, Vansevicius \& Arimoto~2003b; Siebenmorgen \& Krugel~2007; Rowan-Robinson 2012). Early models of this kind did not include the evolution of stellar populations. Silva et al.~(1998) were the first to couple radiative transfer through a dusty ISM and the spectral (and chemical) evolution of stellar populations. 

Modelling the emission from stars and dust consistently in order to get reliable estimates for the main galaxy physical parameters, like stellar mass ($M_{\star}$), star formation rate (SFR) and the average extinction ($A_{V}$), involves solving the radiative transfer equation for idealised but realistic geometrical distributions of stars and dust, as well as taking advantage of the full SED coverage from UV to sub-mm.
The full SED allows the total luminosity of the galaxy to be robustly constrained without relying on extrapolations based on optical data alone.  
While the shape of the optical-UV SED indicates some level of dust extinction, there may be stellar populations (typically the newly born) completely obscured by dust.
The only way to overcome this problem and quantify how much stellar light (and therefore mass) is missing from the optical, is to consider the dust-absorbed galaxy luminosity through the dust re-emitted spectrum in the IR, and model it together with the optical spectrum with a complete radiative transfer method.

With total infrared luminosities between $10^{11}-10^{12}$ L$_{\odot}$ and $\geq 10^{12}$ L$_{\odot}$, respectively, Luminous and Ultraluminous Infrared Galaxies [(U)LIRGs hereafter] are among the most luminous and complex extragalactic objects we can conceive, including all varieties of young and old stellar populations, dust absorption, scattering, grain thermal re-radiation, and AGN emission (Lonsdale et al.~2006).
Although they are quite rare in the local universe, they dominate the cosmic star formation rate and the FIR background at z $>$ 1.
Therefore they are suitable laboratories to study the main physical processes which drive galaxy formation and evolution.
Observations of local ULIRGs have shown that they are dominated by strong interactions and mergers and the fraction of mergers/interactions
among them has been found to be strongly correlated with their IR luminosity, such that lower luminosity LIRGs (L$_{\mathrm{IR}}$ $\lesssim$ $10^{11.5}$ L$_{\odot}$) are ordinary disks while the highest luminosity ULIRGs are advanced stage mergers. 

High-z LIRGs and ULIRGs seem to be, instead, not equivalents of ULIRGs in the nearby universe, but rather upscaled versions of normal galaxies (Elbaz et al. 2011; Nordon et al. 2012; Symeonidis et al. 2009; Rujopakarn et al. 2011). An implication is that high-z ULIRGs could well have gradually evolving SFHs, and be not necessarily associated with merger events as they are at z$\sim$0. 

This paper is the first of a series dedicated to a detailed physical investigation of the spectro-photometric properties of dusty high-redshift (U)LIRGs in an attempt to achieve a deeper understanding of these sources.

Here we concentrate on a small sample of high-$z$ (U)LIRGs with the currently richest suite of photometric and spectroscopic data, available for the first time at redshift $z\sim2$, combining deep \textit{Herschel} imaging with ultra-deep IRS spectra from \textit{Spitzer}. The data is analysed with a state-of-art chemo-spectro-photometric model (GRASIL: Silva et al.1998) including self-consistent treatment of dust absorption and reprocessing based on a full radiative transfer solution. 

In section \S \ref{data} we describe our data sample. The physical modeling performed for this study is presented in section \S \ref{grasilandFadda}. The results concerning our SFH and SFR estimates are discussed in section \S \ref{SFRandSFH}. While the predicted stellar masses and average extinctions for these high-z (U)LIRGs are presented in section \S \ref{massDet}. The conclusions can be found in section \S \ref{Conclusions}.

Our cosmological model assumes H$_{0}$=71 km $s^{-1}$ Mpc$^{-1}$, $\Omega_{\Lambda}$=0.73, $\Omega_{M}$=0.27.

\section{The (U)LIRG data sample}
\label{data}

Our high-$z$ (U)LIRGs have been selected from the sample of 48 IR-luminous galaxies in GOODS-S presented by Fadda et al.~(2010, F10 hereafter). It includes the faintest $24$ $\mu$m sources observed with IRS ($S_{24} \sim 0.15-0.45$ mJy) in the two redshift bins ($0.76-1.05$ and $1.75-2.4$) and samples the major contributors to the cosmic infrared background at the most active epochs. At these redshifts, the $24\ \mu$m fluxes translate to infrared luminosities roughly in the ranges for LIRGs at $z\sim1$ and ULIRGs at $z\sim2$. This sample is therefore crudely luminosity selected (with the $S_{24}$ limits corresponding to different luminosity ranges for the z$\sim1$ LIRGs and z$\sim2$ ULIRGs) and F10 did not apply any other selections. 

These objects have been selected by F10 mostly using the photometric redshifts computed by Caputi et al.~(2006) since only 50\% and 10\% of them had spectroscopic-z in the z$\sim$ 1 and 2 bins, respectively. As emphasized by F10, except for having higher dust obscuration, these galaxies do not have extremely deviant properties in the rest-frame UV/optical compared to galaxies selected at observed optical/near-IR band. Their observed optical/near-IR colors are very similar to those of extremely red galaxy (ERG) populations selected by large area K-band surveys.

In this paper we concentrate on galaxies powered by star formation. Therefore we excluded from our present analysis objects previously classified by F10 as AGN-dominated on the basis of several indicators such as broad and high ionization lines in optical spectra, lack of a $1.6\ \mu$m stellar bump in the SED, X-ray bright sources, low mid-IR $6.2\ \mu$m EW and optical morphology (see also Pozzi et al.~2012). We also excluded the sources falling outside the multiwavelength MUSIC catalogue (Santini et al.~2009), which is our reference for complementary optical/near-IR photometry.
Among the 31 (U)LIRGs fulfilling these requirements, 10 are at $z\sim1$ (all in the nominal LIRG regime) and 21, mostly ULIRGs, are at $z\sim2$. The GOODS-S field was observed in 2010 with both \textit{Herschel} (Pilbrat et al. 2010) SPIRE (Griffin et al.~2010) and PACS (Poglitsch et al.~2010), giving a total of six bands from $70$ to $500$ $\mu$m. SPIRE and PACS data are taken, respectively, from the Herschel Multi-tiered Extragalactic Survey (HerMES: Oliver et al.~2012) and the PACS Evolutionary Probe (PEP: Lutz et al.~2011) programs. Typical noise levels are $\sim 1$ mJy for PACS $70-160$ $\mu$m and $\sim 6$ mJy for SPIRE $250-500$~$\mu$m, including confusion.

\section{Modeling the SEDs of high-z (U)LIRGs}
\label{grasilandFadda}

\begin{table*}[ht]
\centering
\begin{tabular}{|lllll|}
\hline
\hline
Par. & Unit & Description & class & range \\
\hline
\hline
$\nu_{\mathrm{Sch}}$ & $\mathrm{Gyr^{-1}}$ & SF efficiency & SFH: Quiescent & $[0.3-2.0]$\\
$\tau_{\mathrm{inf}}$ & $\mathrm{Gyr}$ & infall timescale & SFH: Quiescent & $[0.01-3.0]$\\
$\%M_{\mathrm{b}}$ & --- & \% of gas mass involved & SFH: burst & $[0.1-15]$\\
    &   &  in the burst at $t_{\mathrm{burst}}$ \footnote{time at which the burst is set on} &   &   \\
$t_{\mathrm{b}}$ & Myr & burst e-folding time & SFH: burst & $[10-80]$\\
  &   &  &  & \\
$f_{\mathrm{mol}}$ & --- & fraction of gas in MC & basic & $[0.05-1.0]$\\
    &  & with respect to the diffuse comp. &     &   \\
$\tau_{1}$ & --- & MC optical depth at 1 $\mu$m & basic & $[5.0-108]$ \\
$t_{\mathrm{esc}}$ & Myr & escape time of newly born & basic & $[1-90]$\\
  &   &  stars from their parent MCs &  &  \\
$\beta$ & --- &  sub-mm dust spectral slope & basic & $[1.5-2.0]$\\
  &    &    &    &    \\
$r_{\mathrm{c}}$ & kpc & core radius & geometrical: King's profile & $[0.01-2.0]$\\
\hline\end{tabular}
\caption{Main GRASIL input parameters and their range of values (see Silva et al.~(1998) for a full description).}
\label{inputPar}
\end{table*}

A physical characterization of the ULIRG phenomenon requires a multi-wavelength approach and a detailed treatment of the effects of dust. The spectro-photometric + radiative transfer code GRASIL (Silva et al. 1998; 2011) satisfies these requirements. It computes the SED of galaxies from far-UV to radio, with a state-of-the-art treatment of dust reprocessing. 
It includes the radiative transfer effects of different dusty environments - star forming molecular clouds (MCs), diffuse dust (cirrus), dusty envelopes around AGB stars, and a geometrical distribution of stars and dust in a bulge-like (King profile) and/or a disk (double exponential) profile. The accounting of the clumping of young stars and dust within the diffuse medium gives rise to the age-dependent attenuation. The dust model consists of grains in thermal equilibrium with the radiation field, and small grains and PAH molecules fluctuating in temperature.

The input star formation histories are computed with CHE$\_$EVO (Silva 1999), a standard chemical evolution code that provides the evolution of the SFR, M$_\mathrm{gas}$ and metallicity, assuming an IMF, a SF law SFR(t)=$\nu_{\mathrm{Sch}}$ $\cdot$ M$_{\mathrm{gas}}$(t)$^{\mathrm{k}}$ + f(t) (i.e. a Schmidt-type SF with efficiency $\nu_{\mathrm{Sch}}$, and a superimposed analytical term to represent transient bursts possibly related to a galactic merger, and an exponential infall of gas ($dM_{\mathrm{inf}}/dt \propto \exp(-t/\tau_{\mathrm{inf}})$). 
A very short infall timescale, $\tau_{\mathrm{inf}}$, can be used to have a so called `close box' chemical evolution model, which ensures that the gas going to form the galaxy is all available at the beginning. This short value of the infall timescale allows us to accrete enough mass in stars of different ages to reproduce the shape of the UV-opt-NIR SED.
In our library of SFHs we consider a wide range of values for $\tau_{\mathrm{inf}}$ from very low (0.01 Gyr) to 2-3 Gyr the latter producing longer phases of gas accretion.
In the following we have adopted k=1, f(t) exponential, and the Chabrier IMF.
Our reference library of SSPs is from Bressan et al.~(1998, 2002), which directly include the effects of dusty envelopes around AGB stars.

We have used a large and fine grid of theoretical SEDs, generated by CHE$\_$EVO+GRASIL. Our SEDs span a wide range of input parameters: star formation history; obscuration times; dust opacities etc (see Tab.~\ref{inputPar}). This grid is first used to explore the parameter space.
Then an object-by-object analysis of the (U)LIRG SEDs was performed in order to refine the fit. We reproduced the observed broadband SEDs of our 31 (U)LIRGs, including a fit to the IRS spectra, which are useful to constrain the physical state of the warm ISM and the PAH intensity.

\begin{figure*}
\includegraphics[width=17.cm, height=17.cm]{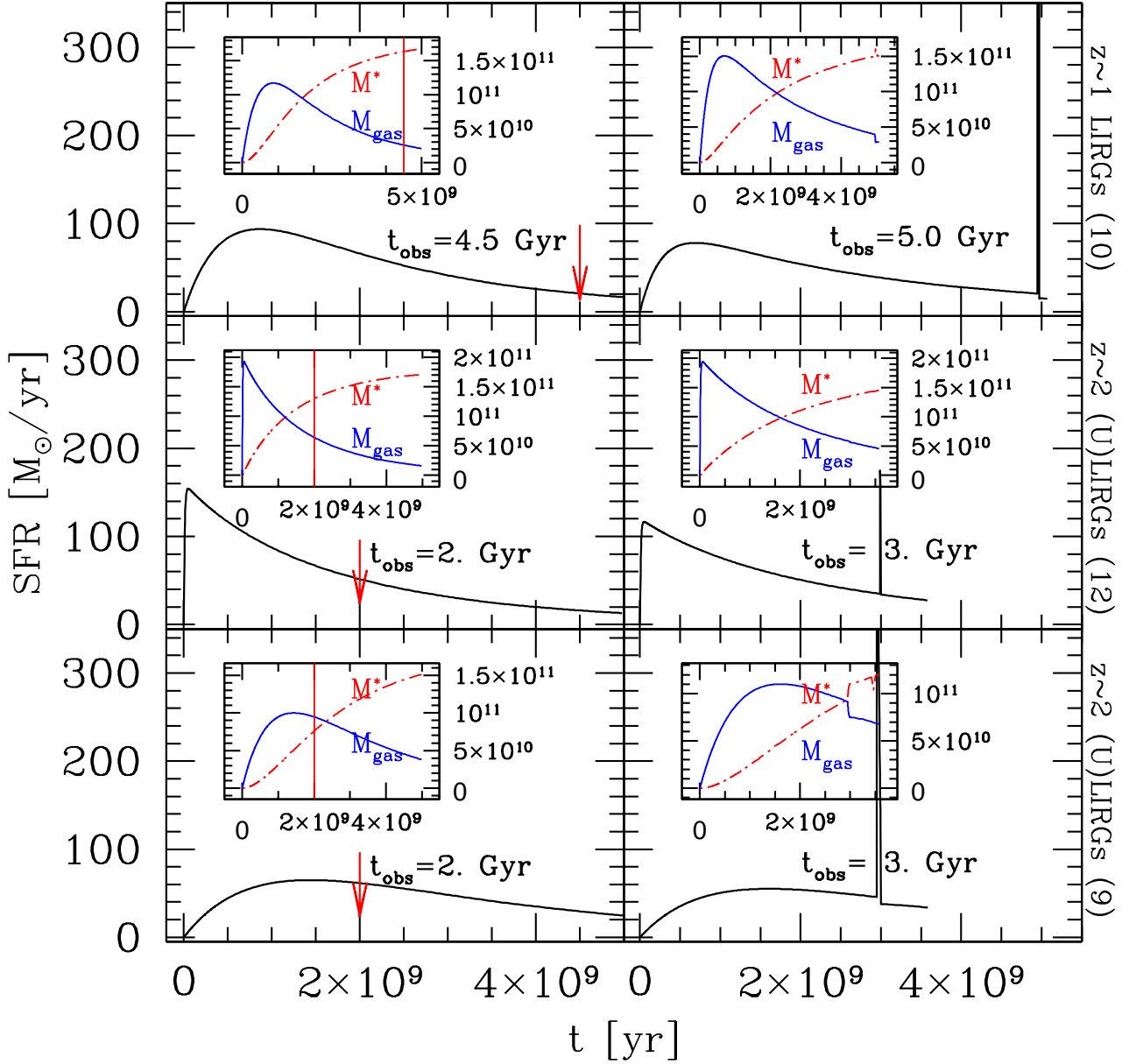}
\caption{Median SFHs of $z\sim1$ (top) and $z\sim2$ (centre and bottom) (U)LIRGs modeled with \emph{continuous} SF (left) or a \emph{starburst} (right). The insets report the time evolution of the mass in living stars and the mass of interstellar gas.
The median values for $\tau_{\mathrm{inf}}$ are 0.5 and 0.28 Gyr in the top-left and top-right panels respectively while they are much shorter and equal to 10 Myr for the centre panels. For the nine z$\sim$2 (U)LIRGs requiring the SFHs represented in the bottom panels the median $\tau_{\mathrm{inf}}$ are 1 Gyr for both.}
\label{SFRtM}
\end{figure*}

Figure \ref{SFRtM} shows the median best-fit SFHs for the 10 $z\sim1$ LIRGs (top) and 21 $z\sim2$ (U)LIRGs (centre and bottom). The insets report the mass in living stars and the mass of interstellar gas. 

\begin{figure*}
\centering
\includegraphics[width=16.5cm,height=14.cm]{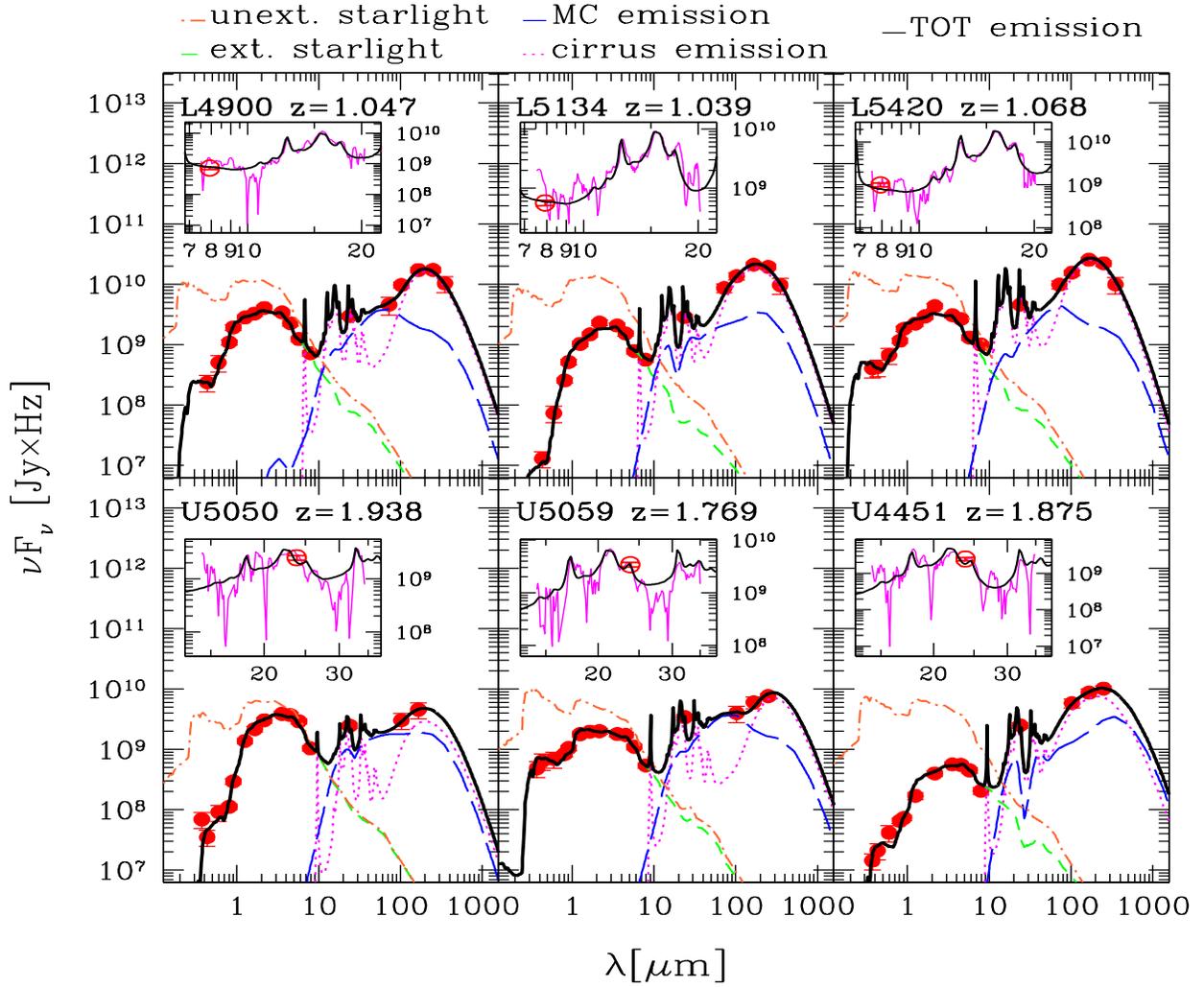}
\caption{Typical GRASIL best-fits (solid black line) to the observed SED (red circles) of three $z\sim1$ (top) and $z\sim2$ (bottom) (U)LIRGs. IRS spectra appear in the inset window (magenta line).
The color-coded lines represent the unextinguished starlight (orange dot-dashed), extinguished starlight (green dashed), cirrus emission (magenta dotted) and MC emission (blue long dashed).}
\label{best-fit}
\end{figure*}

As discussed later, the fitting  process appears to constrain several of the parameters ruling the SFH, in particular $\tau_{\mathrm{inf}}$ and $\nu_{\mathrm{Sch}}$. Small values for $\tau_{\mathrm{inf}}$ and high values for $\nu_{\mathrm{Sch}}$ , i.e. an early fast and efficient SF phase as in the centre panels, were required for 16/31 objects, 4 LIRGs and 12 z$\sim$2 (U)LIRGs, showing the strongest stellar bump in the rest-frame near-IR (as, for example, the ones in Fig.~\ref{best-fit} top-center and bottom-left,right).
Smoother SFHs with longer $\tau_{\mathrm{inf}}$ ($\tau_{\mathrm{inf}}$ $\gtrsim$ 0.2 Gyr and $\tau_{\mathrm{inf}}$ $\gtrsim$ 1 Gyr) are instead required for the z$\sim$1 LIRGs and 9 z$\sim$2 (U)LIRGs shown in the top and bottom panels, respectively. These galaxies present almost `flat' rest-frame NIR bands and higher UV fluxes.  
Figure~\ref{best-fit} summarizes the typical best-fits SEDs of our sample showing as an example three $\mathrm{z}\sim$1 LIRGs (top) and three $\mathrm{z}\sim$2 (U)LIRGs (bottom). The best-fit SEDs of all our galaxies are shown in the Appendix.
\begin{figure*}[ht]
\centerline{
\includegraphics[width=8.6cm]{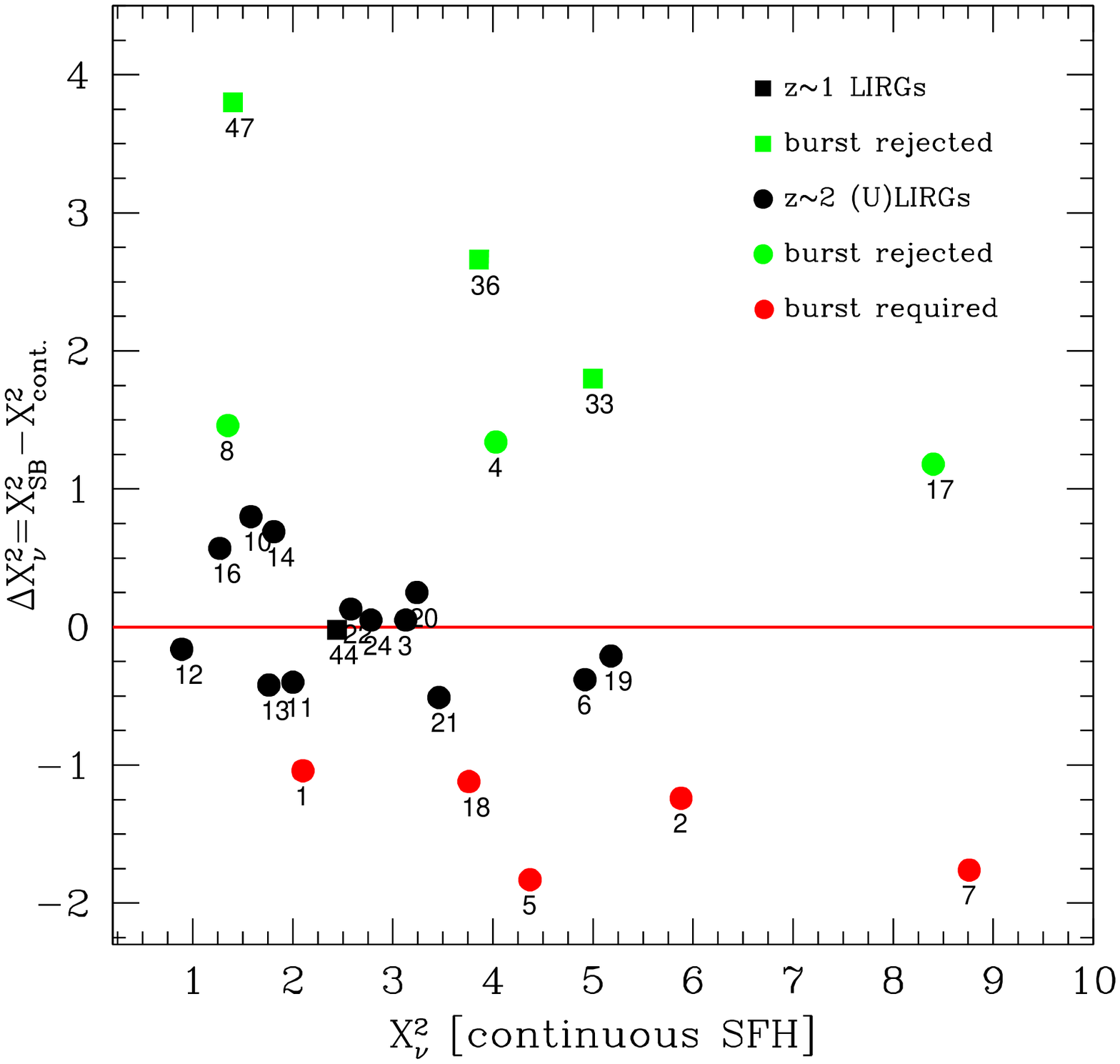} 
\includegraphics[width=8.6cm]{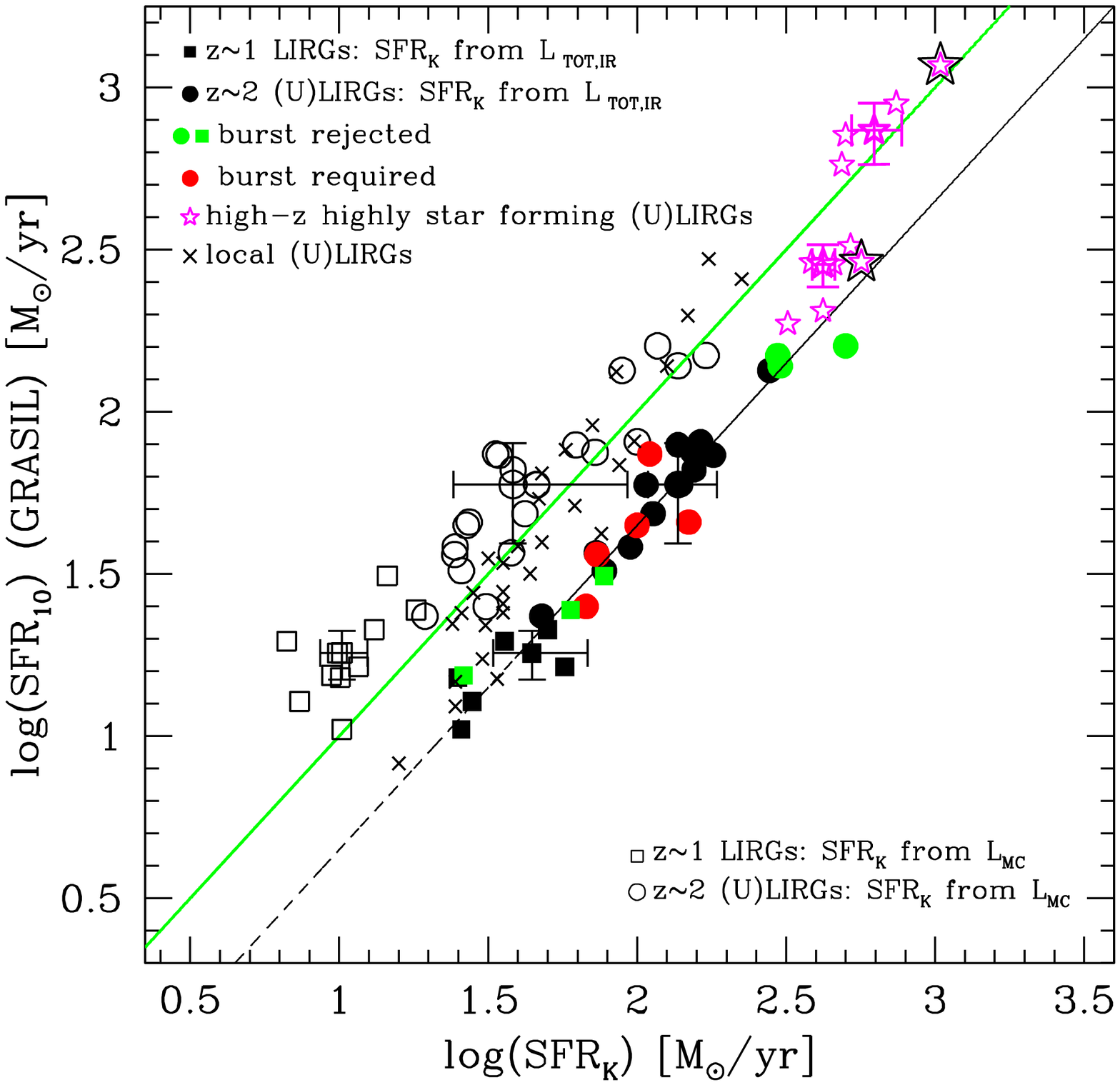}}
\caption{\textbf{Left}: Comparison between the best-fit $\chi^{2}_{\nu}$ relative to the two different prescriptions for the SFH (\emph{continuous} vs \emph{starburst}) for our high-$z$ (U)LIRGs. For 4/10 $z\sim1$ LIRGs (shown as squares) a starburst SFH gave `acceptable' best-fits. For the remaining 6 LIRGs a starbursting SFH was rejected with high-confidence ($\Delta\chi^{2}_{\nu} > 1$). Label numbers indicate the object IDs. \textbf{Right}: Comparison of our GRASIL-estimated $SFR_{10}$ with the SFR derived from the 8 to 1000 $\mu$m luminosity L$_{\mathrm{IR}}$ using the Kennicutt (1998) calibration, $SFR_{\mathrm{K}}$.
Filled and open black squares and circles represent the different $SFR_{\mathrm{K}}$ derived by considering the total $L_{IR}$ (cirrus + star forming MCs) and the $L_{MC}$ (MC only), for our $z\sim1$ and $z\sim2$ (U)LIRGs respectively. Green filled squares and circles are $z\sim1$ and $z\sim2$ (U)LIRGs for which a burst is not required, while red filled circles are $z\sim2$ ULIRGs requiring the presence of a moderate burst from an application of the F-test. Crosses are local (U)LIRGs while magenta stars are high-SFR galaxies discussed by Rodighiero et al.~(2011, R11 hereafter). Datapoints with errorbars are median values with associated semi-interquartile ranges.}
\label{SFRchiH}
\end{figure*}

\section{Evaluating the Star Formation Rates and SFH}
\label{SFRandSFH}

Given the detailed shape of the broadband SED, our physical analysis seems to be able not only to give an estimate of the instantaneous SFR but also to give important hints about the main parameters ruling the source's past SFH, i.e. $\tau_{\mathrm{inf}}$ and $\nu_{\mathrm{Sch}}$ as shown in fig. \ref{SFRtM}.

We have investigated here both SF models with and without a starburst on top of the Schmidt-type part of the SF law (see fig. \ref{SFRtM}). We hereafter refer to these as the \emph{starburst} and \emph{continuous} models, respectively. 
For the majority of our (U)LIRGs, a suitable calibration of the $\tau_{\mathrm{inf}}$ and $\nu_{\mathrm{Sch}}$ allowed us to obtain good fits to the observed SEDs with the continuous models. 
Figure~\ref{SFRchiH} (left) compares the best-fit $\chi^{2}_{\nu}$ relative to these two SFHs considered for our high-$z$ (U)LIRGs.

If we consider first the more extreme cases, i.e., our $z\sim2$ ULIRGs, for 5 out of 21 sources (red filled circles in Fig.~\ref{SFRchiH}), an application of the \textit{F}-test to our $\chi^{2}$ analysis requires the presence of a moderate \emph{starburst}. Even in these five objects, however, the gas mass involved in the SB amounts to only a small fraction of the galactic mass, $\lesssim$ 4\%, and all of them are observed just at the end of the burst event. A peak-phase SF usually produces intense far-IR spectra, with MC emission dominating the MIR and FIR spectrum. Conversely a more continuous SFH generates colder far-IR spectra with {\it cirrus} emission in the sub-mm and larger PAH contributions around 10\ $\mu$m. Two examples are shown in Fig.~\ref{outliers}. This figure shows the typical contributions to the composite SED of a highly star forming and gradually evolving galaxy (left), where the cirrus emission dominates the IR/sub-mm wavelengths, and a pure starburst (right) whose IR fluxes are dominated by MC emission. Both galaxies are taken from the sample of high-z PACS-detected off-main-sequence sources (the so called ``outliers'') of R11 and are modeled with GRASIL. This sample includes all the objects ($\sim$ 50) with sSFR a factor of $\sim$ 8-10 higher than the one of MS galaxies. The $\sim$ 12 galaxies shown as magenta stars in Fig.~\ref{SFRchiH} represent the most extreme cases with the highest SFRs. For the remaining 16 sources of our sample at $z\sim2$, a continuous model gives a perfect account of the data, although without excluding a small starburst contribution (black filled circles in Fig.~\ref{SFRchiH}).

As illustrated in Fig.~\ref{SFRtM}, in most of our objects the only significant effect of introducing a SB event is to slightly increase the age at which the galaxy is observed, $t_{\mathrm{obs}}$, changing from 4.5 to 5 and from 2 to 3 Gyr from left to right, for $z\sim1$ and $z\sim2$ (U)LIRG respectively. The effect of an ongoing SB on the galaxy SED is mimicked, in a continuous SFH, by considering the epoch of observation to be closer to the peak phase of SF.

For the $z\sim1$ LIRGs, all objects are consistent, on the basis of the \textit{F}-test application, with a continuous SFH.

Table~\ref{tabULIRG} shows the main best-fit physical parameters derived from our analysis and their mean values and standard deviations. We report here three different estimates of SFR, SFR$_{\mathrm{K}}$ representing the SFR derived from the 8 to 1000 $\mu$m luminosity $L_{\mathrm{IR}}$ using the Kennicutt~(1998, K98 hereafter) calibration, and our best-fit SFRs averaged over the last $10$ (SFR$_{\mathrm{10}}$) and $100$ (SFR$_{\mathrm{100}}$) Myr.
Fig.~\ref{SFRchiH} (right) shows a comparison of our SFR$_{\mathrm{10}}$ with SFR$_{\mathrm{K}}$. The latter is defined in the limit of complete dust obscuration and dust heating fully dominated by young stars (see K98).

Indeed it assumes that the L$_{\mathrm{bol}}$ of a constant SF lasting 100 Myr is
totally emitted in the IR (K98; Leitherer \& Heckman~1995, LH95 hereafter). For
a constant SF, the L$_{\mathrm{bol}}$ after the first 10 Myr evolves
relatively slowly because the rate of birth and death of the
most massive stars (with lifetimes $\lesssim$ 10 Myr and dominating the L$_{\mathrm{bol}}$) reaches a steady state (see Fig.~2 and 8 of
LH95). The K98 SFR/L$_{\mathrm{FIR}}$ calibration adopts the mean bolometric
luminosity for a 10-100 Myr continuous SF, solar abundance,
Salpeter IMF (which we have rescaled to the Chabrier one) of the starburst synthesis models of LH95,
and assumes that L$_{\mathrm{FIR}}$=L$_{\mathrm{bol}}$. Therefore we consider SFR$_{10}$ as the best indicator of the current SFR to be compared to this calibration.

As apparent in Fig.~\ref{SFRchiH} (right), our inferred SFR$_{10}$ for these high-$z$ (U)LIRGs are systematically lower than those based on the K98 calibration (filled circles and squares). 
\begin{figure}[!h]
\centerline{
\includegraphics[width=8.6cm]{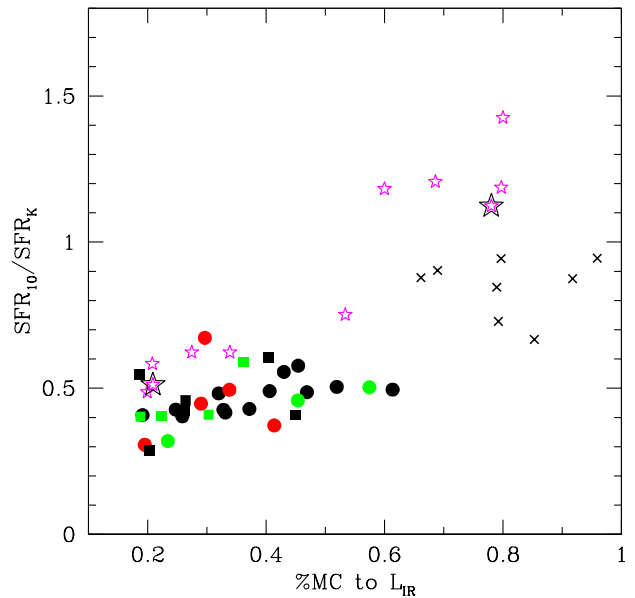}}
\caption{Correlation between the SFR$_{10}$/SFR$_{\mathrm{K}}$ ratio and the fractional contribution of MC emission to the total L$_{\mathrm{IR}}$. Points are the same as in fig.~\ref{SFRchiH}. Here we show the eight local (U)LIRGs for which the fractional contributions of MC and cirrus emission have been already computed by us. Black stars are the two representative cases of R11 off-MS star forming galaxies shown in fig.~\ref{outliers}. We can see here that on average galaxies with higher SFR$_{10}$/SFR$_{\mathrm{K}}$ ratios have total IR luminosity mostly contributed by MC emission with all of our (U)LIRGs lying on the lower relation. Of course the scatter we see in this figure as well as in fig.~\ref{SFRchiH} is due to the wide range of ages of stellar populations.}
\label{ratio}
\end{figure}
\begin{figure*}[ht]
\centerline{
\includegraphics[width=8.6cm]{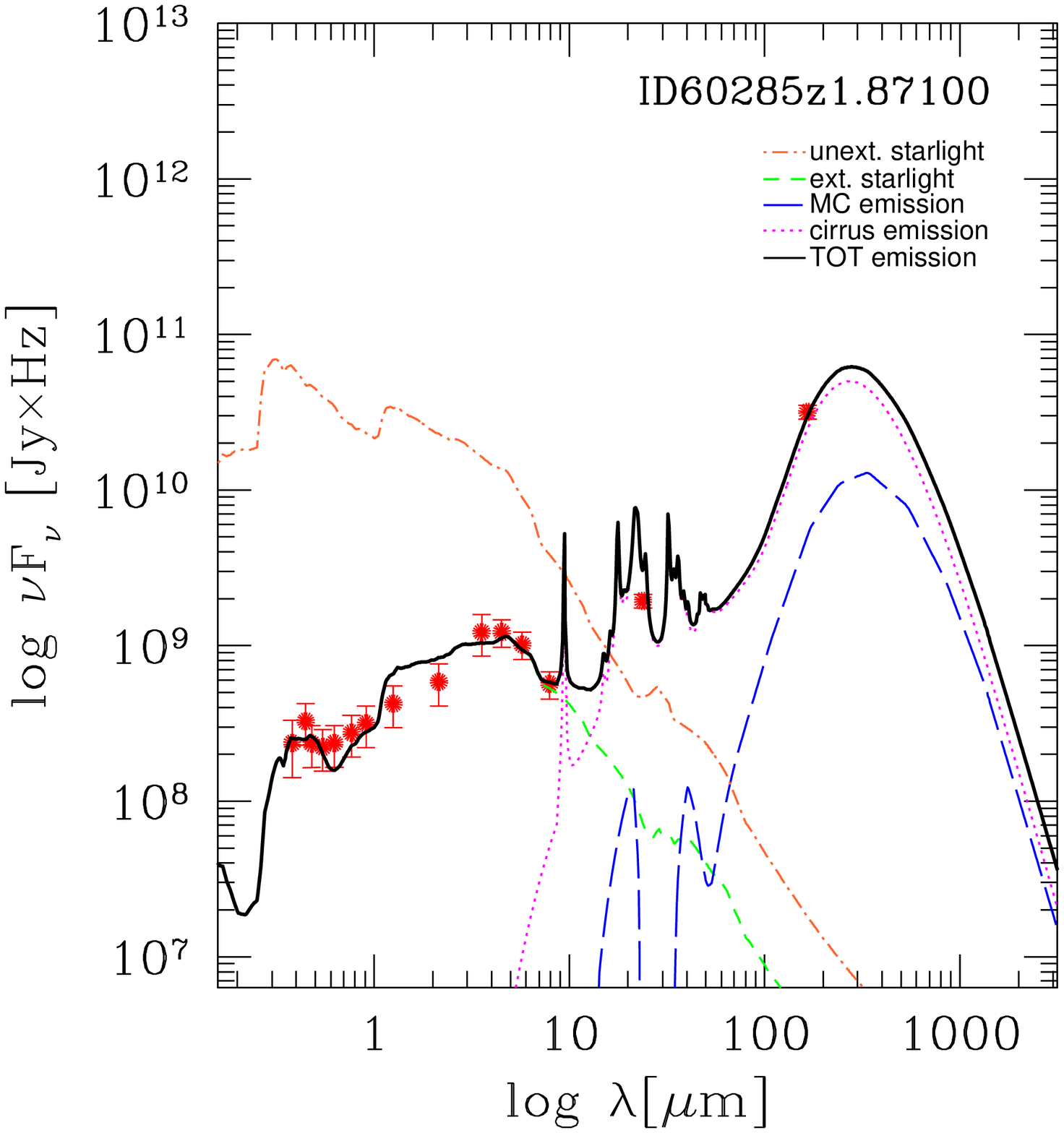}
\includegraphics[width=8.6cm]{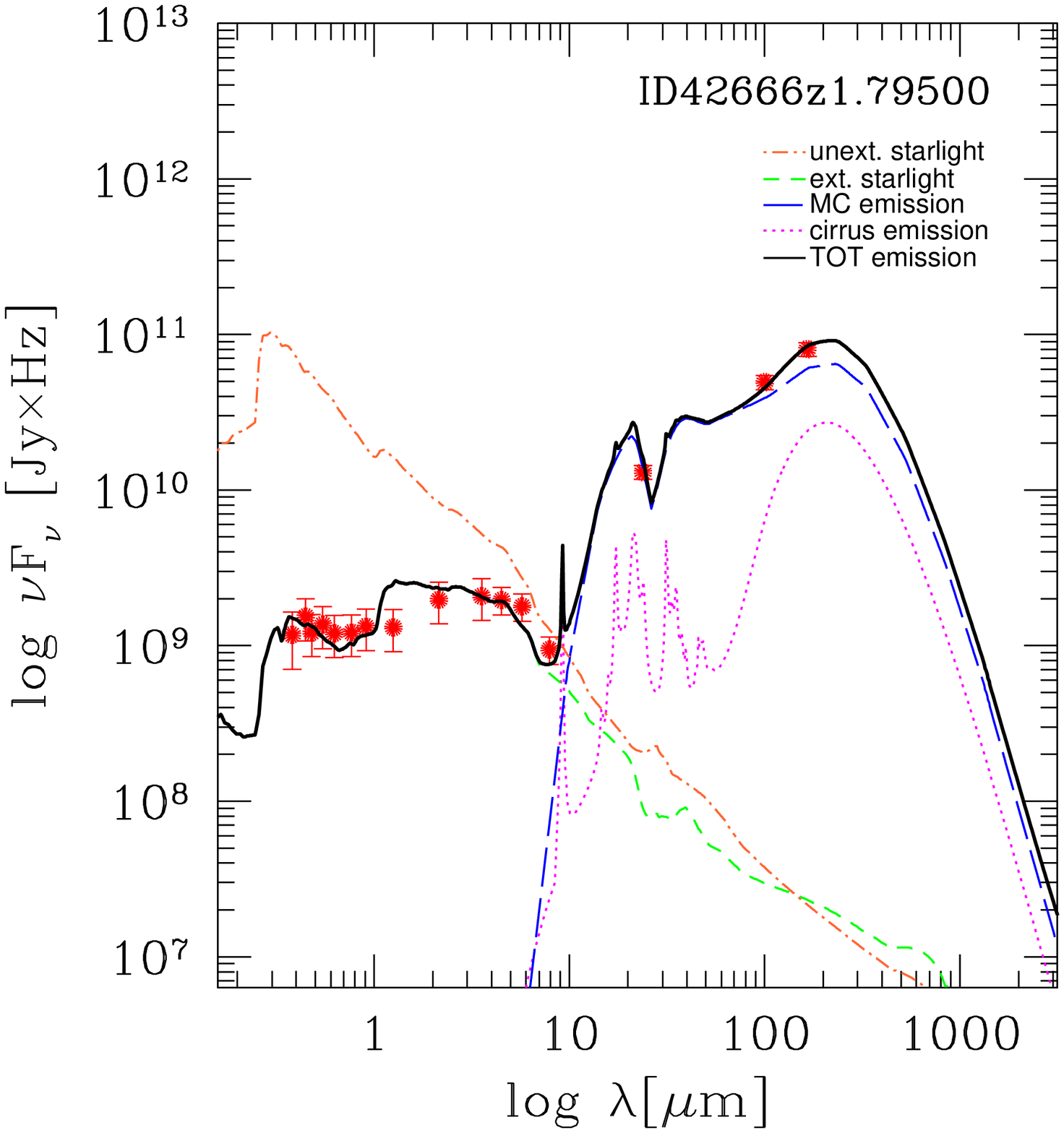}}
\caption{Figure shows two representative cases of the R11 outliers defined in section~\ref{SFRandSFH}. Left: the galaxy is modeled as gradually evolving spheroid and have total L$_{\mathrm{IR}}$ dominated by cirrus emission. Right: the galaxy is dominated by an ongoing burst of star formation in MCs. While the galaxy on the right has SFR in agreement with the one estimated from L$_{\mathrm{IR}}$ using the Kennicutt calibration, the one on the left follows our calibration.}
\label{outliers}
\end{figure*}
This is due to the significant contribution of cirrus emission to the total L$_{\mathrm{IR}}$ ($\sim 73\%$ and $\sim 66\%$ for z$\sim1$ and z$\sim2$ (U)LIRGs respectively) whose heating source includes already evolved stellar populations (ages older than t$_{\mathrm{esc}}$). As cautioned by K98, if all of this energy is ascribed, through this calibration, to the recent SF, the SFR is overestimated. Instead, given the characteristics of the K98 calibration, a fairer comparison between SFR$_{10}$ and SFR$_{\mathrm{K}}$, when an ongoing starburst is not clearly dominating, is to assign to the latter the L$_{\mathrm{IR}}$ by MC only (empty cirles and squares in Fig \ref{SFRchiH} (right)). In this way most of the points spread around and above the K98 line. The spread has to be ascribed to a range of different and non
constant SFR and SFH, and t$_{\mathrm{esc}}$ (ranging between a minimum of 3 Myr and a maximum of 90 Myr, with median values corresponding to 14 and 6 Myr for z$\sim$1 and z$\sim$2 (U)LIRGs respectively), i.e. to a spread of stellar ages contributing to L$_{\mathrm{IR}}$. Averaging out our results for LIRGs and ULIRGs we find the following calibration between the total IR luminosity and SFR:
\begin{equation}
SFR [M_{\odot}/yr] \simeq (5.8 \pm 0.4)\times 10^{-11}\ L_{\mathrm{IR}}/L_{\odot},
\end{equation}
approximately a factor 1.7--2.5 different from the classical Kennicutt relation, both scaled to a Chabrier IMF.

In Fig.~\ref{SFRchiH} (right) we also plot the SFR$_{10}$ vs SFR$_{\mathrm{K}}$ (from total L$_{\mathrm{IR}}$) derived with the same physical analysis, for a sample of local (U)LIRGs (crosses) and some of the most highly star forming objects of Rodighiero et al.~(2011) (magenta stars).

The galaxies shown in this plot (excluding the open points which are the same galaxies as the filled ones, but considering only the MC contribution) seem to define a broad and continuous distribution contained between two correlation lines: an upper envelope dominated by objects with an ongoing burst of SF in MCs as for several local (U)LIRGs and peak-phase SBs, and a lower one mostly populated by secularly evolving galaxies and late starbursts (as the 5/21 objects of our sample discussed above), with substantial contributions of cirrus emission to the total L$_{\mathrm{IR}}$.

Indeed the relative contributions of MC emission to the total L$_{\mathrm{IR}}$ for these sources are shown in Figure~\ref{ratio} as a function of the SFR$_{10}$/SFR$_{\mathrm{K}}$ ratio: galaxies falling in the lower envelope, with low values of SFR$_{10}$/SFR$_{\mathrm{K}}\simeq 0.5$, have small contributions by MC, while those with
\landscape
\begin{table}[tbp!]
\tabcolsep=0.7mm
\begin{tabular}{|c|c|c|c|c|c|c|c|c|c|c|}
\hline
  \multicolumn{1}{|c|}{ID} &
  \multicolumn{1}{c|}{z} &
  \multicolumn{1}{c|}{$L_{\mathrm{IR}}$} &
  \multicolumn{1}{c|}{log$M^{\star}_{\mathrm{GRASIL}}$}&
  \multicolumn{1}{c|}{log$M^{\star}_{\mathrm{HYPERZ}}$} &
  \multicolumn{1}{c|}{log$M_{\mathrm{dust}}$} &
  \multicolumn{1}{c|}{Av} &
  \multicolumn{1}{c|}{Av} &
  \multicolumn{1}{c|}{log$SFR_{\mathrm{K}}$} &
  \multicolumn{1}{c|}{log$SFR_{\mathrm{10}}$} &
  \multicolumn{1}{c|}{log$SFR_{\mathrm{100}}$} \\
 &  &[L/L$_{\odot}$] & [M$_{\odot}$] & [M$_{\odot}$] & [M$_{\odot}$] & GRASIL & HYPERZ & [M$_{\odot}/\mathrm{yr}$] & [M$_{\odot}/\mathrm{yr}$] & [M$_{\odot}/\mathrm{yr}$]\\
\hline
 28-L4177 & 0.842 & 2.50E11 & 10.94 & 10.95 & 8.28 & 0.93 & 0.20 & 1.40 & 1.18 & 1.19\\
 29-L4419 & 0.974 & 2.27E11 & 11.26 & 11.27 & 8.13 & 0.73 & 1.20 & 1.36 & 1.06 & 1.07\\
 30-L4900 & 1.047 & 5.00E11 & 11.22 & 11.06 & 8.51 & 1.56 & 1.00 & 1.70 & 1.33 & 1.34\\
 31-L5134 & 1.039 & 5.71E11 & 11.32 & 10.96 & 8.25 & 2.76 & 1.40 & 1.76 & 1.21 & 1.23\\
 33-L5420 & 1.068 & 7.73E11 & 11.22 & 10.94 & 8.93 & 1.88 & 1.40 & 1.89 & 1.49 & 1.50\\
 35-L5630 & 0.997 & 3.59E11 & 10.85 & 10.76 & 8.6 & 1.03 & 0.40 & 1.55 & 1.29 & 1.30\\
 36-L5659 & 1.044 & 5.98E11 & 11.21 & 10.82 & 8.69 & 2.75 & 1.40 & 1.78 & 1.39 & 1.40\\
 37-L5876 & 0.971 & 2.61E11 & 11.22 & 11.19 & 8.43 & 0.72 & 0.40 & 1.42 & 1.19 & 1.20\\
 44-L13958 & 0.891 & 2.79E11 & 10.82 & 10.77 & 8.0 & 1.32 & 0.20 & 1.45 & 1.11 & 1.12\\
 47-L15906 & 0.976 & 4.43E11 & 10.9 & 10.48 & 8.36 & 2.25 & 1.20 & 1.65 & 1.26 & 1.27\\	
  \hline
  \hline	
 $\bar{x}$ $\pm$ $\sigma$  & 0.98 $\pm$ 0.07  & (4.26$\pm$1.73)$10^{11}$ & 11.10 $\pm$ 0.18 & 10.92 $\pm$ 0.22 & 8.42 $\pm$ 0.26 & 1.59 $\pm$ 0.74 & 0.88 $\pm$ 0.49 & 1.59 $\pm$ 0.18 & 1.25 $\pm$ 0.12 & 1.26 $\pm$ 0.12\\
\hline\end{tabular}
\ \\ \
\tabcolsep=0.7mm
\begin{tabular}{|c|c|c|c|c|c|c|c|c|c|c|}
\hline
  \multicolumn{1}{|c|}{ID} &
  \multicolumn{1}{c|}{z} &
  \multicolumn{1}{c|}{$L_{\mathrm{IR}}$} &
  \multicolumn{1}{c|}{log$M^{\star}_{\mathrm{GRASIL}}$}&
  \multicolumn{1}{c|}{log$M^{\star}_{\mathrm{HYPERZ}}$} &
  \multicolumn{1}{c|}{log$M_{\mathrm{dust}}$} &
  \multicolumn{1}{c|}{Av} &
  \multicolumn{1}{c|}{Av} &
  \multicolumn{1}{c|}{log$SFR_{\mathrm{K}}$} &
  \multicolumn{1}{c|}{log$SFR_{\mathrm{10}}$} &
  \multicolumn{1}{c|}{log$SFR_{\mathrm{100}}$} \\
 &  &[L/L$_{\odot}$] & [M$_{\odot}$] & [M$_{\odot}$] & [M$_{\odot}$] & GRASIL & HYPERZ & [M$_{\odot}/\mathrm{yr}$] & [M$_{\odot}/\mathrm{yr}$] & [M$_{\odot}/\mathrm{yr}$]\\
\hline
1-U428 & 1.783 & 1.13E12 & 11.39 & 10.89 & 8.91 & 2.42 & 1.80 & 2.05 & 1.69 & 1.70\\
2-U4367 & 1.624 & 7.52E11 & 11.49 & 11.42 & 8.4 & 1.81 & 1.60 & 1.88 & 1.47 & 1.49\\
3-U4451 & 1.875 & 1.40E12 & 11.31 & 10.63 & 8.95 & 3.11 & 1.80 & 2.15 & 1.78 & 1.79\\
4-U4499 & 1.956 & 3.02E12 & 11.69 & 11.11 & 9.49 & 3.28 & 1.40 & 2.48 & 2.14 & 2.15\\
5-U4631 & 1.841 & 9.21E11 & 11.2 & 10.76 & 8.9 & 2.05 & 1.20 & 1.96 & 1.65 & 1.66\\
6-U4639 & 2.112 & 1.54E12 & 11.08 & 10.86 & 9.18 & 1.39 & 0.80 & 2.19 & 1.87 & 1.88\\
7-U4642 & 1.898 & 7.22E11 & 11.1 & 10.73 & 8.8 & 2.12 & 1.40 & 1.86 & 1.57 & 1.59\\
8-U4812 & 1.93 & 5.00E12 & 11.61 & 11.02 & 9.39 & 3.76 & 2.00 & 2.70 & 2.20 & 2.21\\
10-U4958 & 2.118 & 1.63E12 & 11.12 & 11.00 & 9.21 & 1.9 & 0.60 & 2.21 & 1.91 & 1.92\\
11-U5050 & 1.938 & 7.28E11 & 11.54 & 11.68 & 8.45 & 0.97 & 1.60 & 1.86 & 1.57 & 1.58\\
12-U5059 & 1.769 & 1.07E12 & 11.11 & 11.04 & 9.08 & 1.12 & 1.00 & 2.03 & 1.77 & 1.78\\
13-U5150 & 1.898 & 1.80E12 & 11.17 & 10.62 & 9.09 & 2.87 & 1.40 & 2.25 & 1.87 & 1.87\\
14-U5152 & 1.794 & 1.13E12 & 11.23 & 10.62 & 8.87 & 3.24 & 2.00 & 2.05 & 1.68 & 1.70\\
16-U5632 & 2.016 & 2.78E12 & 11.26 & 11.00 & 9.18 & 1.88 & 0.80 & 2.44 & 2.13 & 2.14\\
17-U5652 & 1.618 & 2.96E12 & 11.58 & 11.14 & 9.4 & 3.68 & 2.60 & 2.47 & 2.18 & 2.19\\
18-U5775 & 1.897 & 1.40E12 & 11.0 & 10.54 & 8.93 & 2.67 & 1.00 & 2.15 & 1.88 & 1.89\\
19-U5795 & 1.703 & 7.76E11 & 10.9 & 10.59 & 9.07 & 2.45 & 0.60 & 1.89 & 1.51 & 1.52\\
20-U5801 & 1.841 & 4.78E11 & 10.97 & 10.71 & 8.61 & 2.54 & 2.00 & 1.68 & 1.37 & 1.38\\
21-U5805 & 2.073 & 1.55E12 & 11.23 & 10.46 & 9.18 & 3.66 & 1.20 & 2.19 & 1.82 & 1.83\\
22-U5829 & 1.742 & 9.48E11 & 11.52 & 11.22 & 8.63 & 2.34 & 1.20 & 1.98 & 1.58 & 1.60\\
24-U16526 & 1.749 & 1.37E12 & 10.35 & 10.24 & 9.13 & 2.6 & 0.80 & 2.14 & 1.90 & 1.88\\
  \hline
  \hline
   $\bar{x}$ $\pm$ $\sigma$  & 1.86 $\pm$ 0.14  & (1.58$\pm$1.04)$10^{12}$ & 11.23 $\pm$ 0.29 & 10.87 $\pm$ 0.33 & 9.00 $\pm$ 0.29 & 2.47 $\pm$ 0.79 & 1.37 $\pm$ 0.52 & 2.12 $\pm$ 0.24 & 1.79 $\pm$ 0.23 & 1.80 $\pm$ 0.23\\
\hline\end{tabular}
\caption{Estimated and average values of the main physical parameters, $M_{\star}$, $A_{V}$ and $SFR$, derived from the GRASIL best-fits to the 31 (U)LIRGs, compared with values based on HYPERZ.}
\label{tabULIRG}
\end{table}
\endlandscape
higher SFR$_{10}$/SFR$_{\mathrm{K}}$ values are dominated by MC emission.

An illustration of this is reported in Fig.~\ref{outliers} for the two PACS-selected sources of R11: the one dominated by MC corresponds to the starred object in the upper figure, while that one dominated by cirrus emission falls in the lower part and shows minimal MC contribution. All these aspects will be further discussed in a forthcoming paper extending the sample to all the outliers analysed in R11 and all the local (U)LIRGs of Vega et al.~(2008).

A bimodality in the star formation efficiency (SFE=SFR/M$_{\mathrm{gas}}$) and Kennicutt-Schmidt relation
($\Sigma_{\mathrm{SFR}} \propto \Sigma^{\mathrm{N}}_{\mathrm{H_{2}}}$) of local and high-$z$ mergers and non- or weakly-interacting star forming galaxies based on CO measurements, has been recently suggested by Genzel et al.~(2010) and Daddi et al.~(2010).
Our upper and lower envelopes in Fig. 3, within which all our sources are contained, may
correspond to these two sequences. Further investigation with larger and unbiased samples of galaxies will help in this sense. The latter would be explained by the fact that, while locally galaxies with IR luminosities exceeding $10^{12}$ L$_{\odot}$ are predominantly associated with merging events (e.g. Dasyra et al. 2006), at high-z these high-IR luminosities may be achieved mostly via cold gas accretion (Powell et al. 2011). Our study confirms that high-z (U)LIRGs are more likely up-scaled versions of normal galaxies rather than equivalents of local (U)LIRGs in terms of the mode of SF (e.g. Symeonidis et al. 2009, Rujopakarn et al. 2011).

With our current assumptions (see~\S~\ref{grasilandFadda}), all our mid-IR selected galaxies appear to include massive populations of old ($>1$\ Gyr) stars and, at the same time, to host a moderate ongoing activity of SF (SFR$_{10}$ $\leq$ 100\ M$_{\odot}$/yr, cf. Fig.~\ref{SFRchiH} (right)). The bulk of stars appear to have been formed a few Gyr before the observation in essentially all cases. Average estimates can be inferred from Fig~\ref{SFRtM}: for the z$\sim$ 2 (U)LIRGs having the SFH shown in the central panel of Fig.~\ref{SFRtM}, about 66-80\% of the stellar mass has formed after $\sim$ 1 Gyr from the beginning of their star formation activity, proportionally higher at higher SF efficiency. For the z$\sim$2 (U)LIRGs presenting, instead, more regular SFH, as those shown in the bottom panel of Fig. \ref{SFRchiH}, about 30-43\% of stellar mass is formed after 1 Gyr. Finally for the LIRGs, on average, $\sim$ 28\% of the stellar mass has already been formed after the first 1 Gyr.

\section{Stellar mass determinations }
\label{massDet}

Another key physical parameter of high-redshift galaxies is the stellar mass M$_{\star}$. Our best-fit estimates
appear in Table~\ref{tabULIRG} where we compare our best-fit M$_{\star}$ to those derived by F10.
They fit BC03 models to the optical-to-8 $\mu$m SED with the HYPERZ code (Bolzonella et al.~2000)
allowing different SFHs, from a constant star formation to an exponentially declining SF with different e-folding times ($\tau$-model), and assuming the Calzetti et al. (2000) attenuation law to account for dust extinction effects. The so called $\tau$-model is essentially an analytical approximation of the Schmidt's law $\psi(t) \propto\exp(-t/\tau)$, with $\tau$=1/$\nu_{\mathrm{Sch}}$.
As shown in Tab.~\ref{tabULIRG}, GRASIL M$_{\star}$ are systematically larger than those  
based on HYPERZ. This discrepancy is higher for the ULIRGs at $z\sim2$, for which it reaches a factor of $\sim$ $6$ in M$_{\star}$ (median factor 2.5), while it is lower for the $z\sim1$ LIRGs whose median factor is $1.4$. 

\begin{figure*}
\centerline{
\includegraphics[width=9.cm]{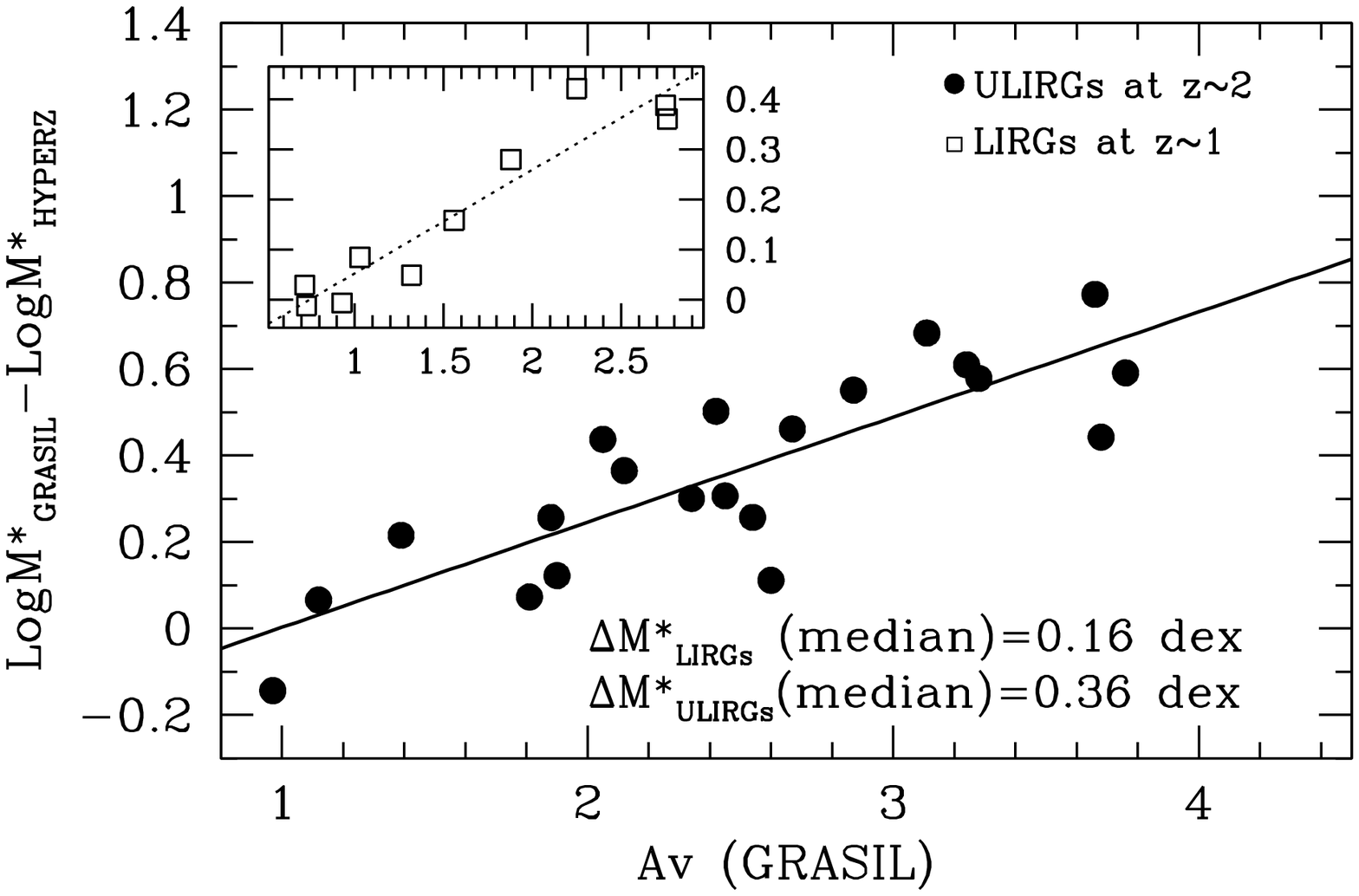}
\includegraphics[width=9.cm]{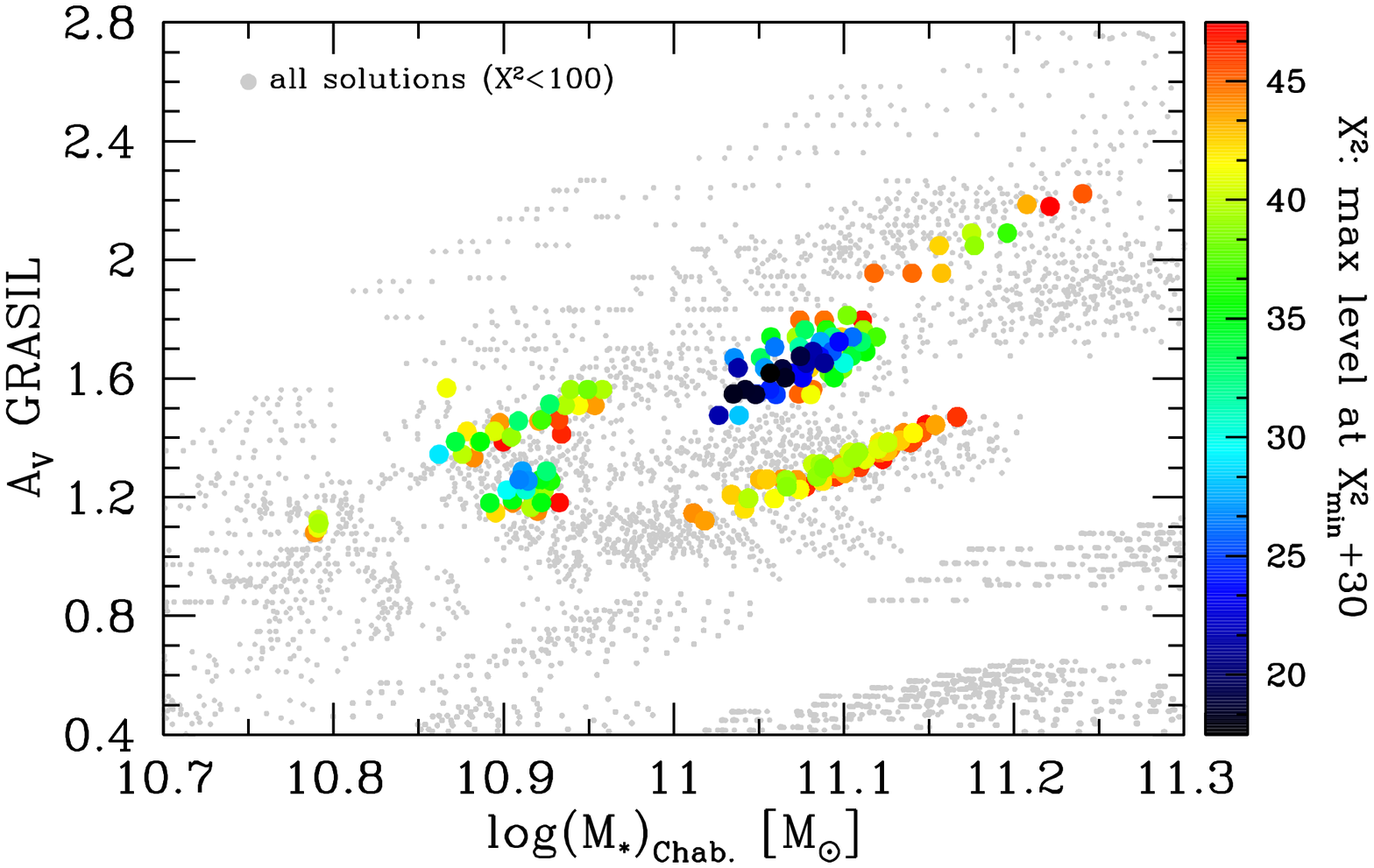}}
\centerline{
\includegraphics[width=8.cm,height=6.5cm]{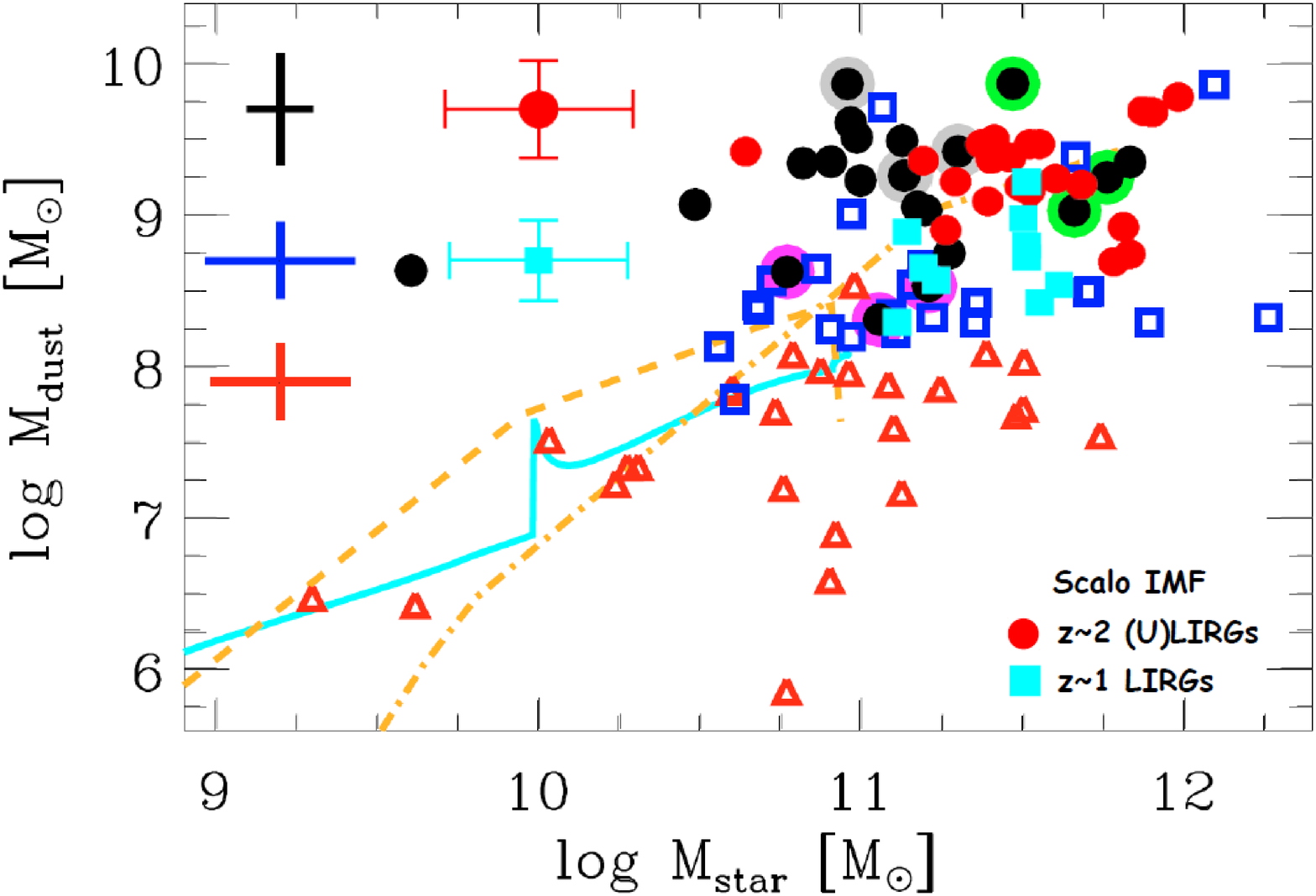}}
\caption{\textbf{Top-Left}: Correlation between the difference in stellar mass estimates based on the GRASIL and HYPERZ codes and the average extinction $A_{V}$. In the inset the LIRGs values are shown. 
\textbf{Top-Right}: $\chi^{2}$ values for spectral models of a typical z$\sim2$ ULIRG are colour-coded ($\chi^{2}$ increasing from dark to light colours) as a function of the average extinction and the stellar mass (see Tab.~\ref{tabULIRG}). Here we show all the solutions with $\chi^{2}$ $\leq$ 100 (gray dots) and  with $\chi^{2}$ within $\chi^{2}_{min}$+30 (colored points).
\textbf{Bottom}: M$_{\star}$ vs M$_{2\mathrm{dust}}$ (from Santini et al.~(2010)). Blue squares are local ULIRGs, red triangles refer to local spirals and black circles correspond to high-$z$ SMGs. Red filled circles and cyan filled squares represent, respectively, our z$\sim2$ and z$\sim1$ (U)LIRGs whose stellar and dust masses have been rescaled to the Scalo IMF adopted by Santini et al. Median 1$\sigma$ error bars are shown on the left. Solid cyan and dashed (dot-dashed) orange lines are the predictions of Calura et al.~(2008) model for spirals and proto-ellipticals with mass of 10$^{11}$ (10$^{12}$) M$_{\odot}$.}
\label{extinction}
\end{figure*}

\subsection{Origin of the stellar mass discrepancy}
\begin{figure*}[ht]
\centerline{
\includegraphics[width=8.6cm]{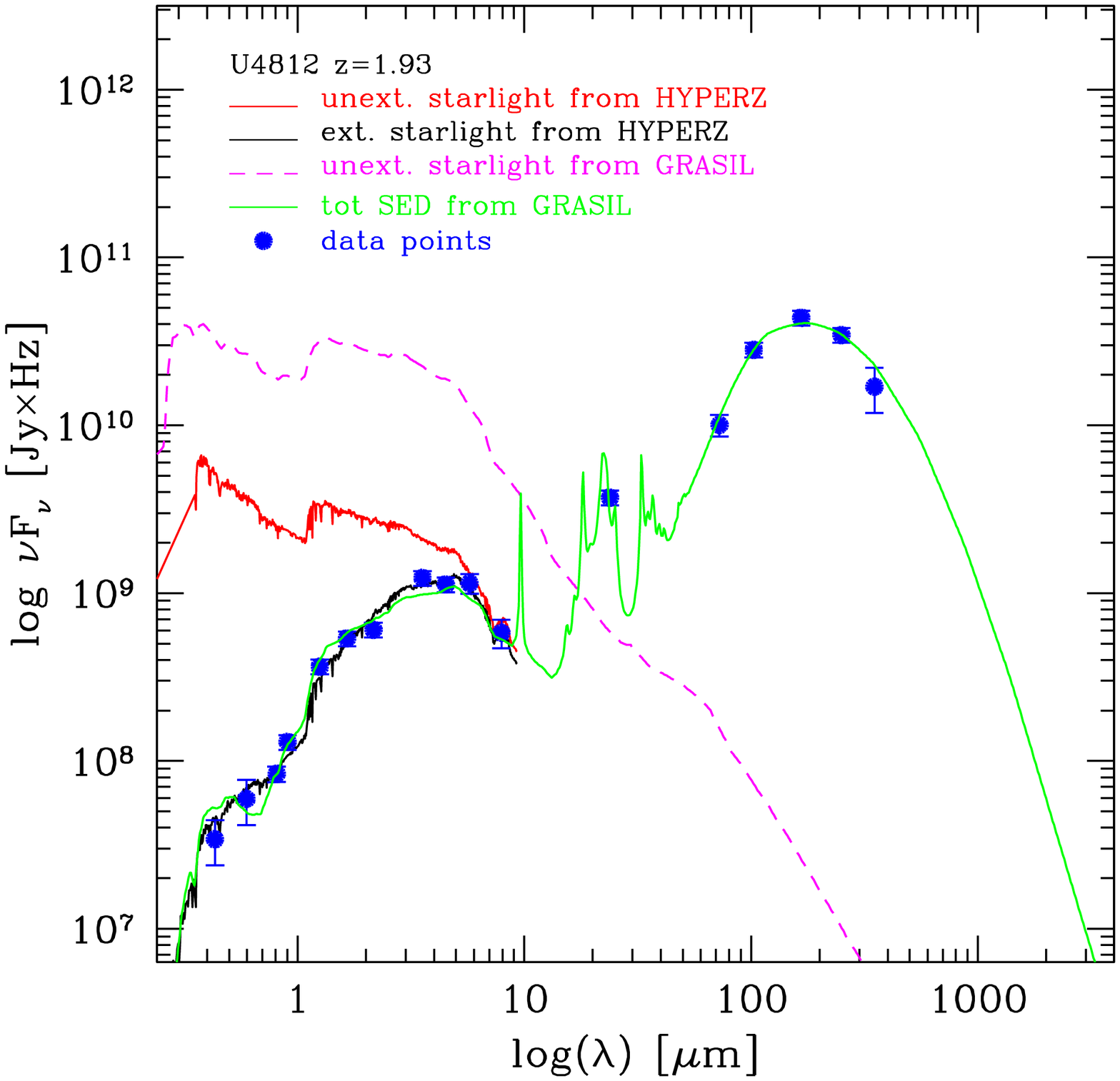}
\includegraphics[width=8.6cm]{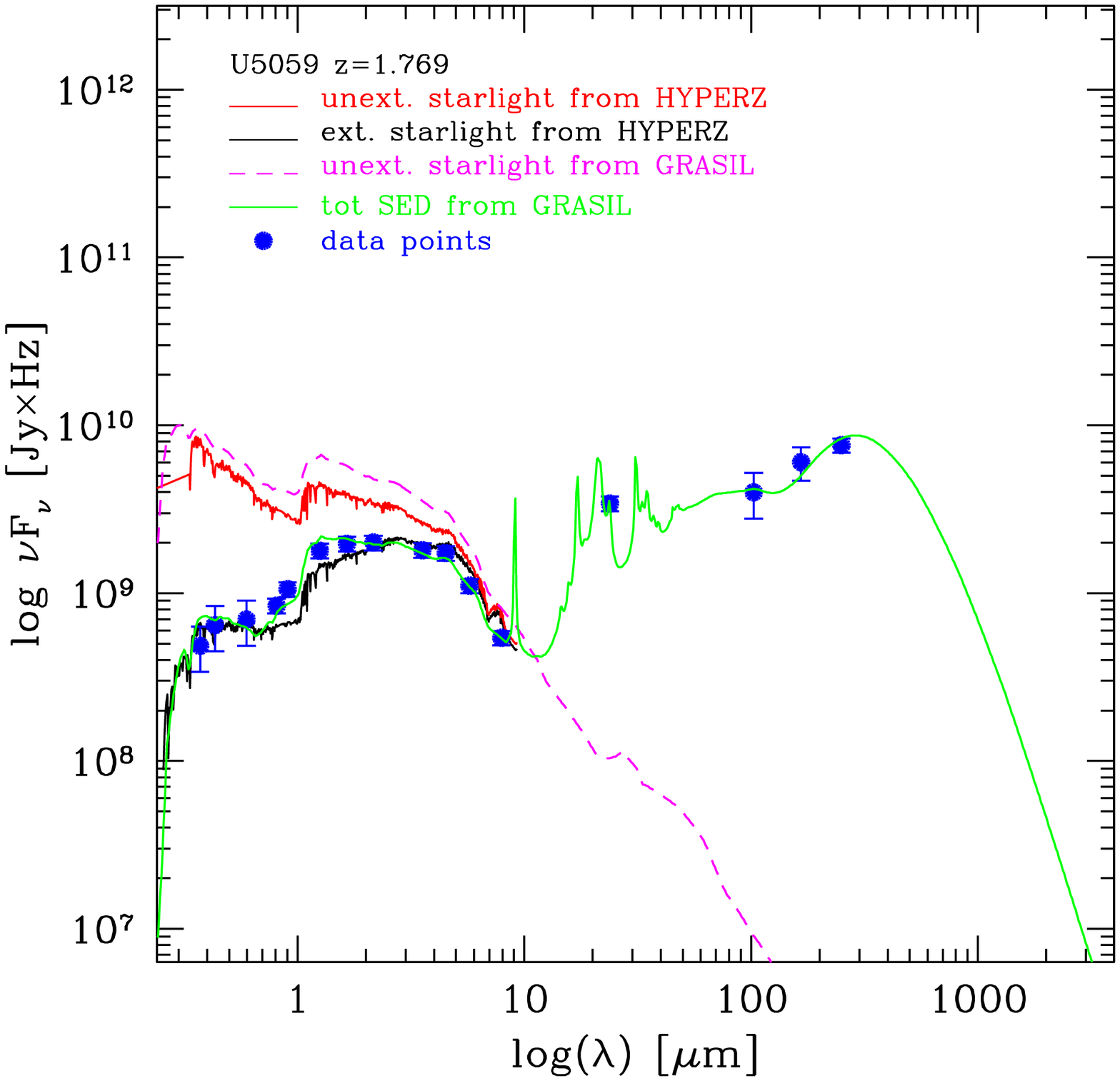}}
\caption{\textbf{Left}:The L$_{\mathrm{bol}}$ estimated from the UV to FIR fit by GRASIL is almost a factor of 4-5 higher than the L$_{\mathrm{bol}}$ given by the optical only fit, at least for those cases were we found large differences in stellar mass estimates. For this galaxy we estimate a M$_{\star}$ $\sim$ 4 times higher.  
\textbf{Right}: Here the effect is smaller. The average extinctions are similar and also the bolometric luminosities. For details see tab.~\ref{tabULIRG}.}
\label{balance}
\end{figure*}
Michalowski et al.~(2012, M12 hereafter) found the adopted specific SFH (double component versus $\tau$-model) to be the major factor affecting the derived stellar masses of high-$z$ and highly star forming SMGs, accounting for a factor of $\sim$ 2.5 in M$_{\star}$. We have investigated the possible origin of the larger differences in our M$_{\star}$ estimates by considering different models for the SFH, continuous vs starburst. In both cases we found similar M$_{\star}$ discrepancies with respect to the HYPERZ ones.
As discussed in detail in M12, the different evoutionary models may also contribute to a factor 2.5 difference in M$_{\star}$ estimates when a $\tau$-model is considered. We tested this hypothesis by modeling our (U)LIRGs with similar SFHs (tau-models), age, metallicity and $A_{\mathrm{V}}$ plus Calzetti attenuation law and by fitting the optical data alone as in the HYPERZ code.
By adopting similar prescriptions we found our M$_{\star}$ to be in full agreement, (i.e. within $10\%$) with the HYPERZ estimates.
Taking into account the differences in the specific parameters ruling the SFH (e.g.~$\tau$) and on the SSP libraries used (BC03 and GRASIL SSPs rely on the same Padova isochrones but the latter includes dusty envelopes around AGB stars) we can conclude that the different SFH (i.e. different $\tau$) and SSPs cannot account for the larger discrepancies, we find among the more obscured ULIRGs at $z\sim2$; for these the dominant factor is the dust exinction. Of course the effect of the different SFH on the stellar mass estimates becomes larger when comparing $\tau$-models with SFH characterized by recent (last 50 Myr) burst of star formation, as shown by M12. All these galaxies are, however, normal star forming galaxies and the only five objects requiring the burst on top of the Schimdt-type SF are all observed as late-SBs.

\subsection{Extinction as the main source of mass discrepancy }

Using the GRASIL best-fits, we estimate the average total extinction in the rest-frame $V$-band, $A_{\mathrm{V}}$, from the ratio of the extinguished ($L_{\mathrm{V}}$) to the unextinguished starlight ($L_{\mathrm{V}}^{0}$): 
\begin{equation}
 \mathrm{A}_{\mathrm{V}} = -2.5 log(L_{\mathrm{V}}/L_{\mathrm{V}}^{0})
\end{equation}
The average A$_{\mathrm{V}}$ values, reported in Table \ref{tabULIRG}, are systematically higher for the $z\sim2$ than for the $z\sim1$ objects.  These differences are traceable to the different evolutionary stages at which the two classes of sources are observed.
Our best-fit A$_{\mathrm{V}}$ are also systematically higher than those obtained with HYPERZ using the Calzetti approximation in combination with the foreground screen of dust, expecially for the $z\sim2$ ULIRGs for which the median A$_{V}$ is $2.45$ and $1.40$ for the two cases, respectively.

As shown in fig.~\ref{extinction} (top-left) we find, for our (U)LIRGs, a tight correlation ($r\sim 0.86$ for $z\sim2$ (U)LIRGs and $r\sim0.96$ for $z\sim1$ LIRGs) between the difference in the M$_{\star}$ estimates based on GRASIL and HYPERZ solutions and the average extinction A$_{\mathrm{V}}$, with the difference increasing as a function of the extinction A$_{V}$. This clarifies why for ULIRGs, with higher A$_{V}$ on average, we see the largest stellar mass discrepancies. 
The stellar mass missed by fitting optical data alone is hidden in dust. In concurrence with our results, Wuyts et al. (2009) also found that the mass underestimate is more severe during the dusty, peak SF phase.

It is worth noting that the availability of the full wavelength coverage also plays a crucial role in constraining the models. In fact the first basic constraint that is provided by the availability of UV-to-sub-mm data is the total L$_{\mathrm{bol}}$ of the galaxy, which otherwise should be guessed. This is easier when the optical contains most of the total energy. The shape of the optical-UV SED indicates some level of dust, but there may be stellar populations totally obscured by dust (typically the newly born), and even less young stellar populations may be partly extinguished. Without a constraint on the optical-UV extinction, i.e. on how much is re-emitted in the IR, it is difficult to quantify how much stellar light (therefore mass) is missing from the optical. Two examples are given in Fig.~\ref{balance} which compares the total L$_{\mathrm{bol}}$ inferred from the unextinguished starlight component computed with HYPERZ (in red) to that one computed with GRASIL (magenta long-dashed line) and to the total L$_{\mathrm{IR}}$ inferred from the GRASIL best-fits. The two objects shown in this figure represent two extreme cases, the one on the left corresponding to a large discrepancy in stellar mass ($\sim$ a factor of 4) and the other one on the right characterized by the lowest stellar mass discrepancy ($\sim$ a factor of 1.17). Of course, due to energy balance, the energy of the pure stellar SED must be equal to that of the processed SED which fit the data. Therefore, the IR part of the SED is a further fundamental constraint when it contains an important fraction of the energy emitted by stellar populations.

Figure~\ref{extinction} (bottom) shows that our stellar and dust masses are consistent with published estimates for high-$z$ SMGs at comparable redshifts. Compared to local spirals and local (U)LIRGs our $z\sim2$ (U)LIRGs have higher dust-to-stellar mass ratios, by a factor of $\sim$ 30 and 6, respectively. Similar results were found by Santini et al.~2010 for the high-$z$ SMGs. The M$_{\mathrm{dust}}$/M$_{\star}$ of z$\sim1$ LIRGs are, instead, similar to those of local (U)LIRGs. M$_{\star}$ and M$_{\mathrm{dust}}$ of galaxies in this work appear also to be in agreement with the predictions of the Calura et al.~(2008) chemical evolution models for proto-ellipticals. The distribution of our sources in this plot is also similar to that found by Rowan-Robinson et al.~2010 for the Hermes Lockman sample. These specific aspects will be dealt with in detail in one of the forthcoming papers of the series.  

The top-right panel of Fig.~\ref{extinction} illustrates the possible degeneracy in the model solutions, between the average extinction and stellar mass, color-coded by the value of $\chi^{2}$, for a typical $z\sim2$ ULIRG. 
As we see, among the many solutions considered, acceptable best-fits, within $\chi^{2}_{\mathrm{min}}$+30, are clearly identified in the parameter space and not much degeneracy is apparent. In particular A$_{\mathrm{V}}$ and M$_{\star}$ seem to be well constrained within $\Delta {\mathrm{A}_{\mathrm{V}}}\sim$ $0.1$ and $\Delta {\mathrm{M}_{\star}}\sim$ $0.04$ dex. These values refer to the case of fixed SFH, that is to say fixed $\tau_{\mathrm{inf}}$ and $\nu_{\mathrm{Sch}}$. If we consider in addition to the many combinations of GRASIL parameters also the different combinations of parameters ruling the SFH, the scatter around the best-fit solution increases to $\Delta {\mathrm{A}_{\mathrm{V}}}\sim$ $0.2$ and $\Delta {\mathrm{M}_{\star}}\sim$ $0.15$ dex, well within the typical uncertainties for this kind of analysis. This is likely to be a result of the continuous spectral coverage from UV to sub-mm for our sample. Another likely factor is the self-consistent radiative transfer modelling, accounting for the distribution of dust with respect to young and old stars, from which both the age-dependent dust attenuation and dust reprocessing are constrained.

\section{Summary and Conclusions}
\label{Conclusions}

The present work reports on a selection of results from a physical self-consistent study of IR luminous galaxies at $z\sim 1$ and $2$. 
For this study we combined far-infrared photometry from two related deep surveys, PEP and HERMES, carried out with the {\textit{Herschel Space Observatory}}, with deep photometry in the near-IR and mid-IR with the {\textit{Spitzer}} IRAC and MIPS imagers, as well as complete mid-infrared spectroscopic follow-up with the {\textit{Spitzer}}'s infrared spectrograph IRS. 

Our working sample includes 31 Luminous and Ultraluminous Infrared Galaxies selected in GOODS-S with $24$ $\mu$m fluxes between 0.2 and 0.5 mJy. Among them 10 are at z$\sim$1 and 21 are at z$\sim$2. They make the faintest high-redshift galaxies observed spectroscopically with IRS, with an excellent suite of photometric data in all bands from UV to the millimeter, including our key far-IR and sub-mm photometry from {\textit{Herschel}}.

These data have prompted us to investigate the nature and the main physical properties - like stellar mass, bolometric luminosity, star-formation history, extinction, as well as the mass assembly history - for every object.
The novelty of our approach consists in modeling the data with a self-consistent physical code (GRASIL), exploiting a state-of-the-art treatment of dust extinction and reprocessing.

We find that all of our galaxies require massive populations of old ($>1$\ Gyr) stars and, at the same time, host a moderate ongoing activity of SF (SFR$_{10}\leq$ 100\ M$_{\odot}$/yr with SFR$_{10}$ being the SFR averaged over the last 10 Myr). Our detailed analysis appears to give important hints also on the past history of SF with the bulk of stars consistent with having been formed a few Gyr before the observation in essentially all cases. Only five galaxies of the sample seem to require, on the basis of an application of the $F$-test to our $\chi^{2}$ analysis, a recent starburst superimposed on a low-level, secularly evolving star formation history. Even in these five objects, however, the gas mass involved in the SB amounts to only a small fraction of the galactic mass, $\lesssim$ 4\%, and all of them are observed just at the end of the burst event (i.e. late SBs).

We find substantial discrepancies between our results and those based on optical-only SED fitting for the same objects. Our physically consistent best-fit model solutions of the observed SEDs indicate higher extinctions (by $\Delta$A$_{\mathrm{V}}\sim$ 0.81 and 1.14) and higher stellar masses (by $\Delta$Log(M$_{\star})\sim$ 0.16 and 0.36 dex) for z$\sim$1 and z$\sim$2 (U)LIRGs, respectively. The stellar mass difference is larger for the most obscured objects and correlates with the total dust extinction. This is in agreement with the results of Wuyts et al. (2009) who also found that the mass underestimate is more severe during the dusty, peak SF phase.

We also find lower SFRs on average than those computed from the total $L_{\mathrm{IR}}$ using the Kennicutt (1998) calibration due to the significant contribution to the dust heating by intermediate-age stellar populations through `cirrus' emission: only $\sim$27\% and $\sim$34\% of the total $L_{\mathrm{IR}}$ appears to be due to ongoing SF in MCs for $z\sim1$ and $z\sim2$ (U)LIRGs, respectively. 

We have demonstrated with the present study the importance of a self-consistent treatment of dust extinction and reprocessing effects in luminous star-forming galaxies via radiative transfer modelling. In future works many aspects will be further investigated, such as a more complete understanding of the degeneracies in our model solutions, including exploring the effects of also other IMFs. This is of particular relevance
for a further refinement of the galaxy star-formation histories and the estimates of stellar mass functions and SFR functions, that have a significant impact on the studies of galaxy formation and evolution.

\acknowledgments
We acknowledge support from ASI (Herschel Science Contract
I/005/07/0). PACS has been developed by a consortium of institutes
led by MPE (Germany) and including UVIE (Austria); KU
Leuven, CSL, IMEC (Belgium); CEA, LAM(France);MPIA (Germany);
INAF-IFSI/OAA/OAP/OAT, LENS, SISSA (Italy); IAC
(Spain). This development has been supported by the funding agencies
BMVIT (Austria), ESA-PRODEX (Belgium), CEA/CNES
(France), DLR (Germany), ASI/INAF (Italy), and CICYT/MCYT
(Spain). SPIRE has been developed by a consortium of institutes
led by Cardiff Univ. (UK) and including: Univ. Lethbridge
(Canada); NAOC (China); CEA, LAM (France); IFSI, Univ. Padua
(Italy); IAC (Spain); Stockholm Observatory (Sweden); Imperial
College London, RAL, UCL-MSSL, UKATC, Univ. Sussex (UK);
and Caltech, JPL, NHSC, Univ. Colorado (USA). This development
has been supported by national funding agencies: CSA
(Canada); NAOC (China); CEA, CNES, CNRS (France); ASI (Italy); MCINN (Spain); SNSB (Sweden); STFC,UKSA (UK); and
NASA (USA).

\newpage
\appendix

\section{APPENDIX: Best-fit SEDs of the entire sample of high-z (U)LIRGs}
%
\begin{figure*}[ht]
\centering
\includegraphics[width=17.cm,height=19.cm]{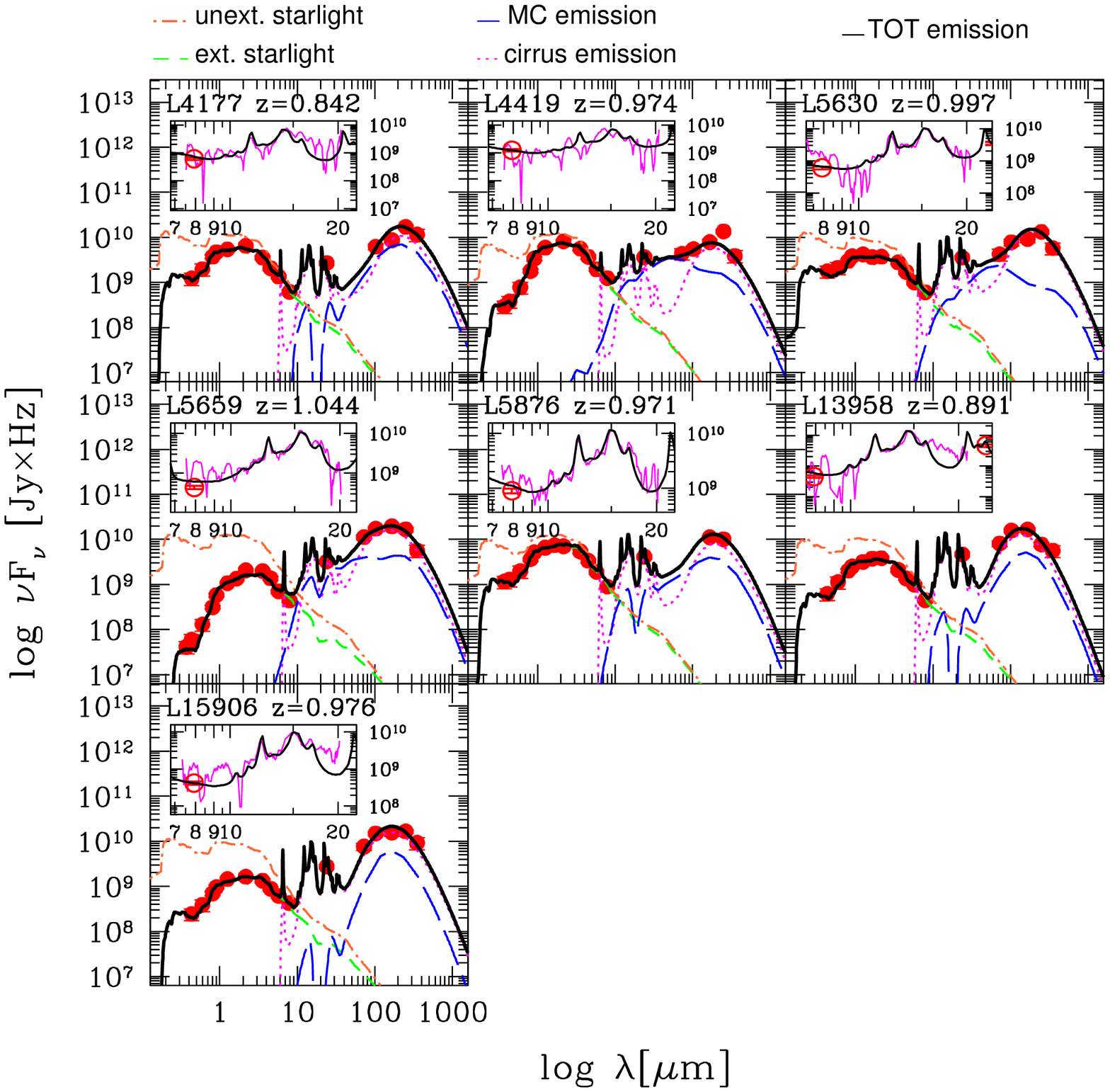}
\caption{GRASIL best-fits (solid black line) to the observed SEDs (red circles) of $z\sim1$ and $z\sim2$ (U)LIRGs. IRS spectra appear in the inset window (magenta line).
The color-coded lines represent the unextinguished starlight (orange dot-dashed), extinguished starlight (green dashed), cirrus emission (magenta dotted) and MC emission (blue long dashed).}
\label{best-fit2}
\end{figure*}
\begin{figure*}
\centering
\includegraphics[width=17.cm,height=19.cm]{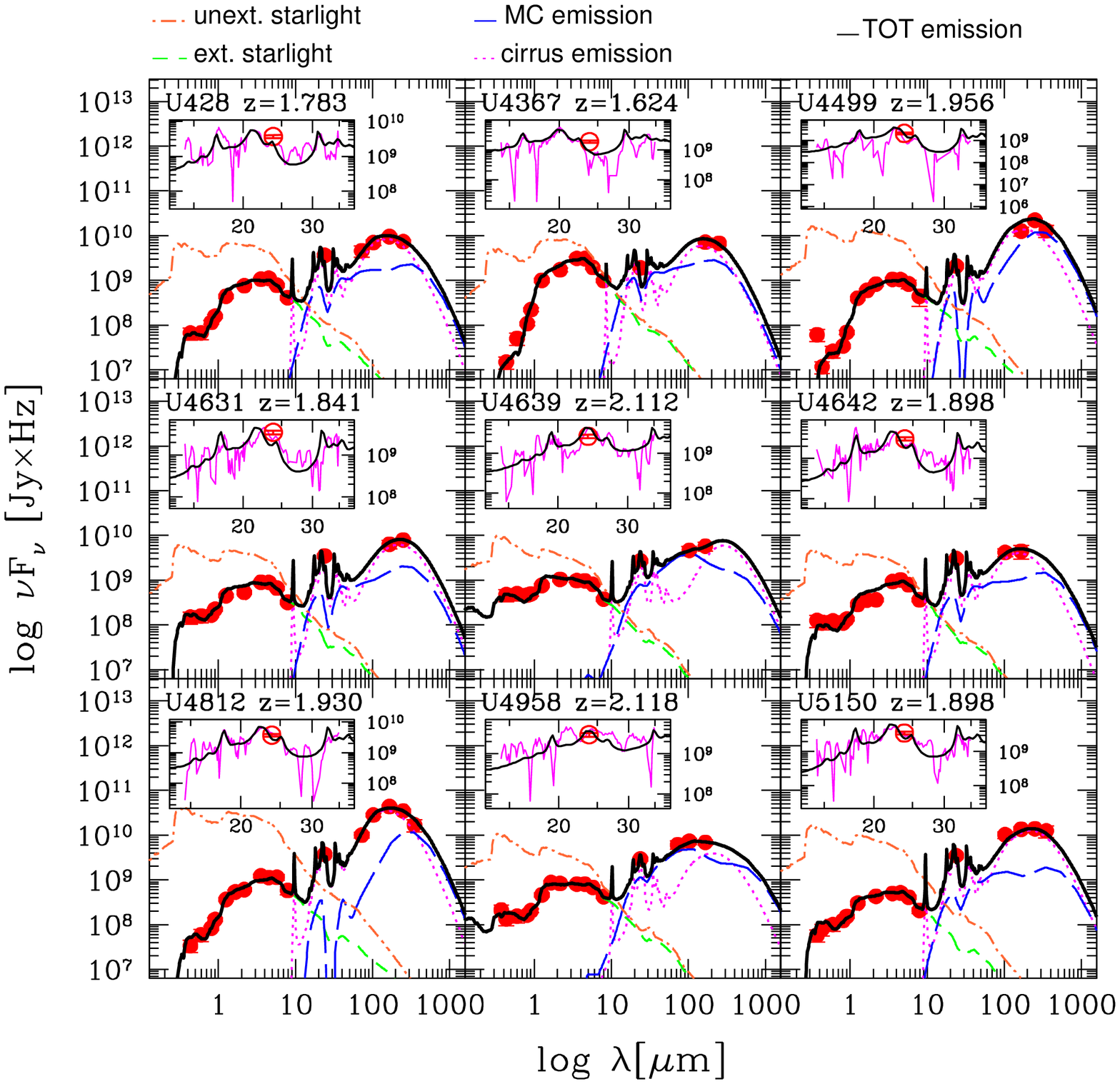}
\caption{same as in fig. \ref{best-fit2}.}
\label{best-fit3}
\end{figure*}

\begin{figure*}
\centering
\includegraphics[width=17.cm,height=19.cm]{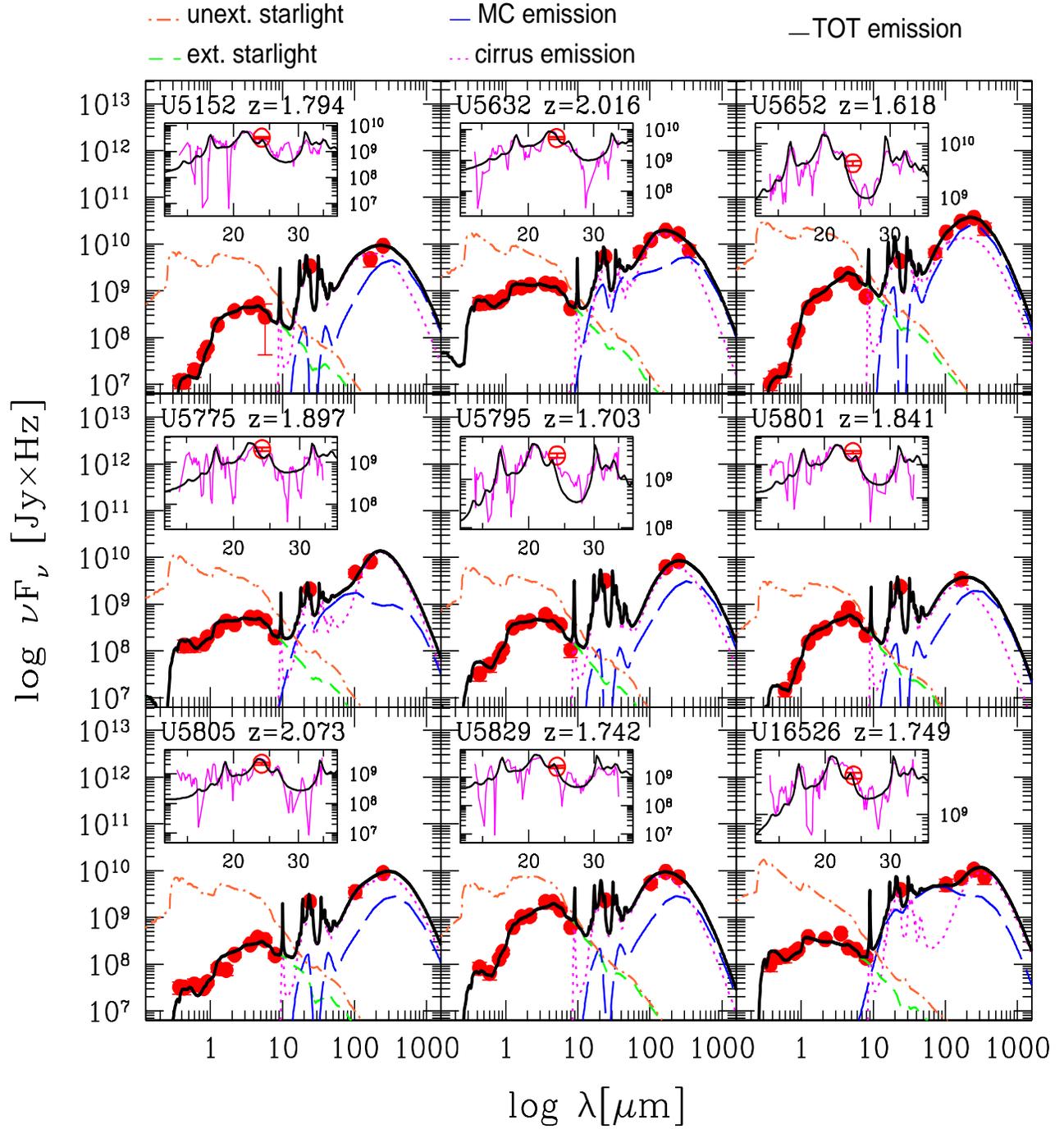}
\caption{same as in fig. \ref{best-fit2}.}
\label{best-fit4}
\end{figure*}

\end{document}